\documentclass[twoside,preprintnumbers,amsmath,amssymb,pacs,nofootinbib]{revtex4}
\usepackage{graphicx}
\usepackage{dcolumn}
\usepackage{bm}
\usepackage{braket}
\usepackage{float}
\usepackage{txfonts}
\usepackage{pslatex}
\usepackage{amssymb}
\usepackage[parfill]{parskip}

\newcommand{\half}{\mbox{\small{$\frac{1}{2}$}}}
\newcommand{\MSbar}{\overline{\mbox{MS}}}

\newcommand{\p}{\partial}
\newcommand{\NA}{N_{\!A}}
\newcommand{\e}{\ensuremath{\mathrm{e}}}

\renewcommand{\d}{\ensuremath{\mathrm{d}}}

\newcommand{\lms}{\Lambda_{\overline{\mbox{\tiny{MS}}}}}
\newcommand{\omu}{\overline{\mu}}

\DeclareMathOperator{\tr}{tr} 
\def\XXint#1#2#3{{\setbox0=\hbox{$#1{#2#3}{\int}$}\vcenter{\hbox{$#2#3$}}\kern-.5\wd0}} % gebruik als \Xint{blabla} om een integraalteken dwars door "blabla" te krijgen

\newcommand{\Nf}{N_{\!f}}

\begin{document}

\title{\textbf{The asymmetry of the dimension 2 gluon condensate: the zero temperature case}}
\author{D.~Dudal$^{a}$, J.~A.~Gracey$^b$, N.~Vandersickel$^a$, D.~Vercauteren$^a$, H.~Verschelde$^a$}\email{david.dudal@ugent.be,gracey@liv.ac.uk,nele.vandersickel@ugent.be,david.vercauteren@ugent.be,henri.verschelde@ugent.be}
 \affiliation{\vskip 0.1cm
                            $^a$ Ghent University, Department of Mathematical Physics and Astronomy \\
                            Krijgslaan 281-S9, B-9000 Gent, Belgium\\\\\vskip 0.1cm
                            $^b$ Theoretical Physics Division, Department of Mathematical Sciences, University of Liverpool\\ P.O. Box 147, Liverpool, L69 3BX, United Kingdom\\\\\vskip 0.1cm
}

\begin{abstract}
\noindent We provide an algebraic study of the local composite operators $A_\mu
A_\nu-\frac{\delta_{\mu\nu}}{d}A_{\kappa}^2$ and $A_\mu^2$, with $d=4$ the spacetime dimension. We prove that these are
separately renormalizable to all orders in the Landau gauge. This corresponds to a renormalizable decomposition of the operator $A_\mu A_\nu$ into its trace and traceless part. We present explicit results for the relevant renormalization group functions to three loop order, accompanied with various tests of these results. We then develop a formalism to determine the zero temperature effective potential for the corresponding condensates, and recover the already known result for $\braket{A_\mu^2}\neq0$, together with \mbox{$\Braket{A_\mu
A_\nu-\frac{\delta_{\mu\nu}}{d}A_{\kappa}^2}=0$}, a nontrivial check that the approach is consistent with Lorentz symmetry. The formalism is such that it is readily generalizable to the finite temperature case, which shall allow a future analytical study of the electric-magnetic symmetry of the $\braket{A_\mu^2}$ condensate, which received strong evidence from recent lattice simulations by Chernodub and Ilgenfritz, who related their results to 3 regions in the Yang-Mills phase diagram.
\end{abstract}
\preprint{LTH-824}\maketitle

\section{Introduction}
The dimension 2 gluon condensate $\braket{A_\mu^2}$ in pure Yang-Mills theory has been proposed in \cite{Gubarev:2000eu,Gubarev:2000nz}, and it has been investigated in different ways since then \cite{Verschelde:2001ia,Dudal:2002pq,Dudal:2003vv,Vercauteren:2007gx,Dudal:2005na,Boucaud:2001st,Furui:2005he,Gubarev:2005it,Browne:2003uv,Andreev:2006vy,RuizArriola:2006gq,Chernodub:2008kf}.

In \cite{Verschelde:2001ia} an analytical framework for studying this condensate has been developed, based on work carried out in the Gross-Neveu model \cite{Verschelde:1995jj}. Different problems had to be overcome. First of all there is the gauge invariance of this condensate. In order to make the operator $A_\mu^2$ gauge invariant, one can take the minimum of its integral over the gauge orbit. Since $\int \d^dx\, A_\mu^U A_\mu^U$, with $U\in SU(N)$, is positive, this minimum will always exist. In a general gauge, however, the minimum is a highly nonlocal and thus hard to handle expression of the gauge field. A minimum is however reached in the Landau gauge ($\p_\mu A_\mu=0$), though, so that working in this gauge reduces the operator to a local expression\footnote{We ignore the Gribov problem here, see also \cite{Dudal:2005na}.}. Secondly adding a source $J$, coupled to $A_\mu^2$, makes the theory nonrenormalizable at the quantum level. To solve this, a term quadratic in the source must be added, which in turn spoils the energy interpretation for the effective action. One way around this is to perform the Legendre inversion, but this is rather cumbersome, especially so with a general, spacetime dependent source. One can also use a Hubbard-Stratonovich transform, which introduces an auxiliary field (whose interpretation is just the condensate) and eliminates the term quadratic in the source. Details can be found in \cite{Verschelde:2001ia}. The result was that the Yang-Mills vacuum favors a finite value for the expectation value of $A_\mu^2$. The precise renormalization details of the procedure proposed in \cite{Verschelde:2001ia} were given in \cite{Dudal:2002pq}.

Recently, Chernodub and Ilgenfritz \cite{Chernodub:2008kf} have considered the asymmetry in the dimension two condensate. They performed lattice simulations, computing the expectation value of the electric-magnetic asymmetry in Landau gauge, which they defined as
\begin{equation}\label{ci}
N_c\Delta = \langle g^2 A_0^2 \rangle - \frac1{d-1} \sum_{i=1}^{d-1} \langle g^2 A_i^2 \rangle\, .
\end{equation}
At zero temperature, this quantity must of course be zero due to Lorentz invariance\footnote{We shall deliberately use the term Lorentz invariance, though we shall be working in Euclidean space throughout this paper.}. Necessarily it cannot diverge as divergences at finite $T$ are the same as for $T=0$, hence this asymmetry is in principle finite and can be computed without renormalization, for all temperatures. The authors of \cite{Chernodub:2008kf} found that the high-temperature behavior of the asymmetry had no surprises, following a power law as can be guessed from general thermodynamic arguments. For the low-temperature behavior, however, one would expect an exponential fall-off with the lowest glueball mass in the exponent, $\Delta\sim e^{-m_{\text{gl}}T}$. Instead they found an exponential with a mass $m$ significantly smaller than $m_\text{gl}$. So far, there is no explanation for this behavior.

The goal of this work is to construct an analytical framework to investigate the electric-magnetic asymmetry studied numerically in \cite{Chernodub:2008kf}, with the aim of shedding more light on the results of that paper. The hope is that something more will be found concerning the light mass scale $m$ influencing the thermal behavior of this condensate. It was also noticed that the behavior of the asymmetry divides the Yang-Mills phase diagram into
three regions in terms of the temperature. Remarkably, these regions seem to coincide with those associated with the condensed,
liquid and gaseous states of the magnetic monopoles, whose dynamics are closely related to (de)confinement, see also \cite{Chernodub:2006gu}. The condensate $\braket{A_\mu^2}$ was already related to monopoles in \cite{Gubarev:2000eu,Gubarev:2000nz}.

Since the computations are rather involved, we will split this work into two papers. In this paper we present the actual formalism, building on \cite{Verschelde:2001ia,Dudal:2002pq}. Despite the fact that the quantity defined in \eqref{ci} does not need renormalization, it is unclear how to study this object directly with an effective potential approach. Any finite temperature effective potential is a generalization of the $T=0$ potential, however the operator defining \eqref{ci} makes little sense at $T=0$ as it is not Lorentz invariant. One could think about separately studying the temporal part, $\braket{A_0^2}$, and the spatial part, $\braket{A_i^2}$, but this does not solve the Lorentz symmetry problem at $T=0$: it is unclear how to couple these operators to the action such that Lorentz symmetry is maintained, while simultaneously allowing for a study of the ensuing effective potential. In addition, we would also like to recover the original results for $\braket{A_\mu^2} = \braket{A_0^2}+\braket{A_i^2}$, i.e. we desire a clean $T\to0$ limit. We do solve these problems by considering the operator $A_\mu A_\nu$. This operator is subsequently split into its trace and its traceless part, yielding $A_\mu^2$ and the spacetime asymmetry, $A_\mu A_\nu-\frac{\delta_{\mu\nu}}{d}A_\kappa^2$. We shall prove that these two operators are renormalizable at $T=0$, and that a finite effective potential can be constructed for these, by introducing sources $J$ and $k_{\mu\nu}$, which allow to add both operators to the action without jeopardizing Lorentz symmetry. We also explicitly compute the effective potential at zero temperature, and we show that the only minimum is the one found already in \cite{Verschelde:2001ia}, with a finite expectation value for $A_\mu^2$, but with $\braket{A_\mu A_\nu-\frac{\delta_{\mu\nu}}{d}A_\kappa^2}=0$ --- as one would expect. This is already a nontrivial result, meaning that there is no dynamical Lorentz violation.  At the end, one can take a suitable linear combination of both condensates to retrieve the asymmetry defined in \eqref{ci}, at any (non)zero temperature. A priori, it is however more natural to study $A_\mu^2$ and $A_\mu A_\nu-\frac{\delta_{\mu\nu}}{d}A_\kappa^2$ than \eqref{ci}. In a future paper we shall then focus on the temperature dependence of both $\braket{A_\mu^2}$ itself and the asymmetry.

Summarizing the content of the current paper, we have given in section II the renormalization analysis of the operator using the algebraic formalism of \cite{Piguet:1995er}. In section III we illustrate that the Hubbard-Stratonovich transformation can be used to eliminate terms quadratic in the action, and we compute the quantities necessary for the calculation of the effective action. In section IV finally the effective action itself is computed, and its minima are searched for. In section V the conclusions are presented. Some more technical calculations are bundled in the appendices.

\section{Algebraic analysis of a renormalizable operator and action}
\subsection{Algebraic study of the renormalizability of the local composite operator $A_{\mu }^{a}A_{\nu }^a$}
\subsubsection{The action}
We begin by recalling the expression of the pure Yang-Mills action in the Landau gauge
\begin{eqnarray}
S &=&S_{YM}+S_{GF}  \label{sym} \\
&=&\frac{1}{4}\int\d^dxF_{\mu \nu }^{a}F^a_{\mu \nu
}+\int\d^dx\left( b^{a}\partial _{\mu }A^a_{\mu }+
\overline{c}^{a}\partial _{\mu }D_{\mu }^{ab}c^{b}\right) \;,
\nonumber
\end{eqnarray}
where
\begin{equation}
D_{\mu }^{ab}\equiv \partial _{\mu }\delta ^{ab}-gf^{abc}A_{\mu }^{c}\;.
\label{cd}
\end{equation}
In order to study the local composite operator (LCO) $A_{\mu}^{a}A^a_{\nu }$, we introduce it
in the action by means of a BRST\ doublet of external sources
$\left( K_{\mu\nu}, \eta_{\mu\nu} \right) $, symmetrical in the
Lorentz indices, in the following way
\begin{eqnarray}  \label{ren}
S_{K} & = & s\int\d^dx\left(   \frac{1}{2}  \eta_{\mu\nu} A_{\mu }^{a}A^{a}_\nu - \frac{1}{2 d}  \eta A_{\mu }^{a}A^{a}_\mu - \frac{\omega}{2} \eta_{\mu\nu} K_{\mu\nu}  + \frac{\omega }{ 2d} \eta K \right) \nonumber\\
&=& \int\d^dx\left( \frac{1}{2} K_{\mu\nu} A_{\mu }^{a}A^{a}_\nu
+\eta_{\mu\nu} A^{a}_\mu \partial_{\nu}c^{a} -    \frac{1}{2d}  K
A_{\mu }^{a}A^{a}_\mu - \frac{1}{d} \eta A^a_{\mu} \partial_{\mu}
c^a - \frac{\omega}{2} K_{\mu\nu} K_{\mu\nu} + \frac{\omega
}{2d}K^{2}  \right) \;,
\end{eqnarray}
where $s$ denotes the BRST\ nilpotent operator acting as
\begin{eqnarray}
sA_{\mu }^{a} &=&-D_{\mu }^{ab}c^{b}\;,  \nonumber \\
sc^{a} &=&\frac{1}{2}gf^{abc}c^{b}c^{c}\;,  \nonumber \\
s\overline{c}^{a} &=&b^{a}\;,  \nonumber \\
sb^{a} &=&0\;,  \nonumber \\
s\eta_{\mu\nu} &=&K_{\mu\nu}\;,  \nonumber \\
sK_{\mu\nu} &=&0\;.  \label{s}
\end{eqnarray}
We have shortened the notation by setting
\begin{eqnarray}
 K_{\mu\mu} &=& K\;,\nonumber\\
  \eta_{\mu\mu} &=& \eta\;.
 \end{eqnarray}
In eq.~(\ref{ren}) we have used the property $\eta_{\mu\nu} = \eta_{\nu \mu}$, hence
\begin{eqnarray}
 \eta_{\mu\nu} A^{a}_\mu s (A^a_\nu) &=&  \eta_{\mu\nu} A^{a}_\mu D_{\nu}^{ab}c^{b} =\eta_{\mu\nu} A^{a}_\mu \partial_{\nu}c^{a}\;.
\end{eqnarray}
As is apparent from expressions (\ref{sym}) and (\ref{ren}), the action $\left( S_{YM}+S_{GF}+S_{K}\right) $ is BRST invariant
\begin{equation}
s\left( S_{YM}+S_{GF}+S_{K}\right) =0\;.  \label{si}
\end{equation}
We notice that we can rewrite the action $S_K$ in terms of
$K_{\mu\nu}-\frac{1}{d}\delta_{\mu\nu}K$, also the vacuum term $-\frac{\omega}{2}K_{\mu\nu}K_{\mu\nu}+\frac{\omega}{2d}K^2$, since
\mbox{$(K_{\mu\nu}-\frac{1}{d}\delta_{\mu\nu}K)^2=K_{\mu\nu}^2-\frac{1}{d}K^2$},
so we might be tempted to immediately introduce a traceless tensor
source $k_{\mu\nu}\equiv K_{\mu\nu}-\frac{1}{d}\delta_{\mu\nu}K $ coupled to $A_\mu A_\nu$. However, as not all
components of $k_{\mu\nu}$ can then be considered as independent due
to the constraint $k_{\mu\mu}=0$, using the derivative w.r.t.~$k_{\mu\nu}$ becomes rather tricky, and hence also writing down
suitable Ward identities. The current parametrization in terms of a
completely general source $K_{\mu\nu}$ is thus much more useful. We draw attention to the fact that we are actually coupling $K_{\mu\nu}$ to the (traceless) operator
\begin{equation}\label{operator}
 \mathcal{O}_{\mu\nu}= \frac{1}{2}A_\mu^a
A_\nu^a -\frac{1}{2}\frac{\delta_{\mu\nu}}{d}A^a_\kappa A_\kappa^a\;,
\end{equation}
meaning that we are considering the renormalization of this
particular operator $\mathcal{O}_{\mu\nu}$. As we shall soon find out, we can write down a
sufficiently powerful set of Ward identities which shall ensure that
$\frac{1}{2}A_\mu^a A_\nu^a -\frac{1}{2}\frac{1}{d}A^a_\kappa A_\kappa^a$ is a
renormalizable operator on its own.

According to the LCO philosophy \cite{Verschelde:1995jj,Verschelde:2001ia}, the dimensionless parameter $\omega$ is needed to account for the divergences present in the vacuum Green function $\left\langle\mathcal{O}(x)\mathcal{O}(y)\right\rangle $,which shall turn out to be proportional to the specific (traceless) combination of $K^{2}$ and $K_{\mu\nu} K_{\mu\nu}$ already written down in \eqref{ren}.

\textbf{Remark.} We use the following definition for the derivative
w.r.t.~a symmetric source $\Lambda_{\mu\nu}$:
\begin{equation}\label{der}
    \frac{\delta \Lambda_{\mu\nu}}{\delta
    \Lambda_{\alpha\beta}}=\frac{1}{2}\left(\delta_{\mu\alpha}\delta_{\nu\beta}+\delta_{\mu\beta}\delta_{\nu\alpha}\right)\;.
\end{equation}

\subsubsection{Ward identities}
In order to translate the BRST\ invariance (\ref{si}) into the corresponding Slavnov-Taylor identity, we introduce two further external sources $K _{\mu }^{a}$, $L^{a}$ coupled to the non-linear BRST variations of $A_{\mu }^{a}$ and $c^{a}$
\begin{equation}
S_{\mathrm{ext}}=\int\d^d x\left( -K ^{a }_{\mu}D_{\mu
}^{ab}c^{b}+\frac{1}{2}gf^{abc}L^{a}{c}^{b}c^{c}\right) \;,
\label{sext}
\end{equation}
with
\begin{equation}
s K _{\mu }^{a}=sL^{a}=0\;.  \label{slo}
\end{equation}
Therefore, the complete action
\begin{equation}
\Sigma =S_{YM}+S_{GF}+S_{K}+S_{\mathrm{ext}}\;,  \label{ca}
\end{equation}
obeys the following identities
\begin{itemize}
\item  The Slavnov-Taylor identity
\begin{equation}
\mathcal{S}(\Sigma )=\int\d^d x\left( \frac{\delta \Sigma }{\delta
A_{\mu
}^{a}}\frac{\delta \Sigma }{\delta K ^a _{\mu }}+\frac{\delta \Sigma }{\delta c^{a}}\frac{\delta \Sigma }{\delta L^{a}}+b^{a}\frac{\delta \Sigma }{\delta \overline{c}^{a}}+K_{\mu\nu} \frac{\delta \Sigma }{\delta \eta_{\mu\nu}} \right)  =0\;. \label{sti}
\end{equation}
\item  The Landau gauge fixing condition
\begin{equation}
\frac{\delta \Sigma }{\delta b^{a}}=\partial _{\mu }A^a_{\mu } \;.
\label{gfb}
\end{equation}
\item  The antighost equation
\begin{equation}
\frac{\delta \Sigma }{\delta \overline{c}^{a}}+\partial _{\mu
}\frac{\delta \Sigma }{\delta K_{\mu }^a}=0\;.  \label{ageq}
\end{equation}
\item  The ghost Ward identity
\begin{equation}
\mathcal{G}^{a}\Sigma =\Delta _{\mathrm{cl}}^{a}\;,
\end{equation}
with
\begin{eqnarray}
\mathcal{G}^{a} &=&\int\d^d x\left( \frac{\delta }{\delta
c^{a}}+gf^{abc}\left( \overline{c}^{b}\frac{\delta }{\delta b^{c}}
\right) \right) \;,
\end{eqnarray}
and
\begin{equation}
\Delta _{\mathrm{cl}}^{a}=g\int\d^d xf^{abc}\left( K_{\mu}^{b}A_{\mu
}^{c}-L^{b}c^{c}\right) \;.
\end{equation}
Notice that the term $\Delta _{\mathrm{cl}}^{a}$, being linear in
the quantum fields $A_{\mu }^{a}$, $c^{a}$, is a classical breaking.
\item Thanks to the specific way we introduced the sources
$K_{\mu\nu}$, $\eta_{\mu\nu}$ and their traces $K$, $\eta$, we also
have
\begin{eqnarray}\label{extraward}
\delta_{\mu\nu}\frac{\delta}{\delta K_{\mu\nu}}\Sigma&=&0\;,\nonumber\\
\delta_{\mu\nu}\frac{\delta}{\delta \eta_{\mu\nu}}\Sigma&=&0\;.
\end{eqnarray}
\end{itemize}
Let us also display, for further use, the quantum numbers of all fields and
sources entering the action $\Sigma $
\begin{equation}\label{fields-table}
\stackrel{}{
\begin{tabular}{|c|c|c|c|c|c|c|c|c|}
\hline & $A_{\mu }$ & $c$ & $\overline{c}$ & $b$ & $\eta_{\mu\nu} $
& $K_{\mu\nu}$ & $K $ & $ L $ \\ \hline dimension & $1$ & $0$ & $2$
& $2$ & $2$ & $2$ & $3$ & $4$ \\ \hline
ghost number & $0$ & $1$ & $-1$ & $0$ & $-1$ & $0$ & $-1$ & $-2$ \\
\hline
\end{tabular}}
\end{equation}

\subsubsection{Algebraic characterization of the most general counterterm}
In order to characterize the most general local counterterm which
can be freely added to all orders of perturbation theory, we perturb
the classical action $\Sigma $ by adding an arbitrary integrated
local polynomial $\Sigma ^{\mathrm{count}}$ in the fields and
external sources of dimension bounded by four and with zero ghost
number. This $\Sigma ^{\mathrm{count}}$ is however restricted due to
the existence of the Ward identities. More precisely, it amounts to
impose the following conditions on $\Sigma ^{\mathrm{count}}$:
\begin{itemize}
\item The linearized Slavnov-Taylor identity
\begin{equation}
\mathcal{B}_{\Sigma }\Sigma ^{\mathrm{count}}=0\;,  \label{b1}
\end{equation}
where $\mathcal{B}_{\Sigma }$
\begin{eqnarray}
&&\mathcal{B}_{\Sigma }=\int\d^d x\left( \frac{\delta \Sigma
}{\delta
A_{\mu }^{a}}\frac{\delta }{\delta K^a _{\mu }}+\frac{\delta \Sigma }{\delta K^a_{\mu }}\frac{\delta }{\delta A_{\mu }^{a}}+\frac{\delta
\Sigma }{\delta c^{a}}\frac{\delta }{\delta L^{a}}+\frac{\delta \Sigma }{
\delta L^{a}}\frac{\delta }{\delta c^{a}}+b^{a}\frac{\delta }{\delta
\overline{c}^{a}}+K_{\mu\nu}\frac{\delta }{\delta \eta_{\mu\nu}
}\right) \;,    \label{lb}
\end{eqnarray}
obeys
\begin{equation}
\mathcal{B}_{\Sigma }\mathcal{B}_{\Sigma }=0\;.  \label{lbn}
\end{equation}
\item  The Landau gauge fixing condition
\begin{equation}
\frac{\delta \Sigma ^{\mathrm{count}} }{\delta b^{a}}= 0\;.
\label{b2}
\end{equation}
\item  The antighost equation
\begin{equation}
\frac{\delta \Sigma ^{\mathrm{count}}  }{\delta \overline{c}^{a}}+\partial _{\mu }\frac{\delta
\Sigma ^{\mathrm{count}} }{\delta K ^a_{\mu }}=0\;.  \label{b3}
\end{equation}
\item The ghost Ward identity
\begin{equation}\label{b4}
\mathcal{G}^{a}\Sigma ^{\mathrm{count}}=0\;.
\end{equation}
\item The additional identities
\begin{eqnarray}\label{extrawardconstraints}
\delta_{\mu\nu}\frac{\delta}{\delta K_{\mu\nu}}\Sigma^{\mathrm{count}}&=&0\;,\nonumber\\
\delta_{\mu\nu}\frac{\delta}{\delta
\eta_{\mu\nu}}\Sigma^{\mathrm{count}}&=&0\;.
\end{eqnarray}
\end{itemize}
Taking into account that $\left( K_{\mu\nu},\eta_{\mu\nu} \right) $
form a BRST doublet, from the general results on the cohomology of
Yang-Mills theories it turns out that the external sources $\left(
K_{\mu\nu},\eta_{\mu\nu} \right) $ can only contribute through terms
which can be expressed as pure $\mathcal{B}_{\Sigma }$-variations.
Henceforth, the invariant local counterterm
$\Sigma^{\mathrm{count}}$ can be parametrized as
\cite{Piguet:1995er}
\begin{equation}
\Sigma ^{\mathrm{count}}=\frac{a_0 }{4}\int\d^d xF_{\mu \nu
}^{a}F^a_{\mu \nu }+\mathcal{B}_{\Sigma }\Delta ^{-1}\;,
\label{cnterm}
\end{equation}
where $a_0 $ is a free parameter and $\Delta ^{-1}$ is the most general local polynomial with dimension $4$ and ghost number $-1$, given by
\begin{eqnarray}
\Delta ^{-1} &=&\int\d^d x\left( a_{1}K _{\mu
}^{a}A_{\mu}^a+a_{2}L^{a}c^{a}+a_{3} \partial _{\mu
}\overline{c}^{a}A_{\mu }^a+\frac{a_{4}
}{2}gf_{abc}\overline{c}^{a}\overline{c}^{b}c^{c}\right.   \nonumber \\
 &&\left. +a_{5}b^{a}\overline{c}^{a}  +a_{6}\frac{\eta_{\mu\nu}}{2}A_{\mu }^aA_{\nu }^{a}+a_{7}\eta A^a_{\mu} A^a_\mu+a_{8} \frac{\omega }{2}\eta_{\mu\nu} K_{\mu\nu} + a_{9} \frac{\omega }{2}\eta K\right) \;,
\end{eqnarray}
with $a_{1},\ldots,a_{9}$ still arbitrary parameters.

From the conditions (\ref{b2}), (\ref{b3}), \eqref{b4} it
consequently follows that
\begin{align}
a_{3} &=a_{1}\;, &a_{4}&=a_{5}=0\;,   & a_2&=0\;,  \label{az}
\end{align}
and from \eqref{extrawardconstraints} we find
\begin{align}
a_{6} &=a_{7} d , &a_{8}&=a_{9} d\;,
\end{align}
and hence $\Delta ^{-1}$ reduces to
\begin{eqnarray}
\Delta ^{-1} &= & \int\d^d x\left( a_{1} \left( K _{\mu
}^{a}A^{a\mu}+  \partial _{\mu }\overline{c}^{a}A^{a\mu } \right)
+a_{6} \left( \frac{\eta_{\mu\nu}}{2}A_{\mu }^aA_{\nu }^{a}-
\frac{1}{d}\eta A^a_{\mu} A^a_\mu \right) +a_{8} \left(
\frac{\omega }{2}\eta_{\mu\nu} K_{\mu\nu} - \frac{1}{d} \frac{\omega
}{2}\eta K \right) \right) \;.
\end{eqnarray}
Therefore, for the final form of the most general possible
counterterm one obtains\footnote{It is formally understood that we
work in dimensional regularization, with $d=4-\epsilon$. We have
left the $d$ in front of the operators instead of writing $4$, as
this is important in order to get the correct finite parts once
calculating in $d$ dimensions.}
\begin{eqnarray}\label{finalcount}
\Sigma ^{\mathrm{count}} &=&a_0 \int\d^d x \frac{1}{4} F_{\mu \nu }^{a}F_{\mu \nu }^a+a_1  \int\d^d x\left( A^a_\mu \frac{\delta S_{YM}}{\delta A^a_\mu} + K^a_\mu \partial_\mu c^a + \partial \overline c^a \partial c^a +  K_{\mu\nu} A_{\mu }^{a}A^{a}_\nu +\eta_{\mu\nu} A^{a}_\mu \partial_{\nu}c^{a} -\frac{1}{d}K A_{\mu }^{a}A^{a}_\mu -\frac{1}{d}\eta A^a_{\mu} \partial_{\mu} c^a  \right) \nonumber\\
&& + a_6 \int\d^d x \left( \frac{1}{2} K_{\mu\nu} A_{\mu
}^{a}A^{a}_\nu + \eta_{\mu\nu} A^{a}_\nu \partial_{\mu}c^{a}  -
\frac{1}{2d}  K A_{\mu }^{a}A^{a}_\mu - \frac{1}{d} \eta A^a_{\mu}
\partial_{\mu} c^a\right)  + a_8 \int\d^d x \left(  \frac{\omega}{2}
K_{\mu\nu} K_{\mu\nu} -   \frac{\omega }{2d}K^{2} \right)\;.
\end{eqnarray}
Finally, it remains to discuss the stability of the classical
action, i.e.~to check whether $\Sigma ^{\mathrm{count}}$ can be
reabsorbed in the classical action $\Sigma $ by means of a
multiplicative renormalization of the coupling constant $g$, the
parameters $\omega$, the fields $\left\{ \phi
=A,c,\overline{c},b\right\} $ and the sources $\left\{ \Phi=K,
K_{\mu\nu},\eta, \eta_{\mu\nu} , L,K \right\} $, namely
\begin{equation}
\Sigma (g,\omega ,\phi ,\Phi )+\vartheta\Sigma
^{\mathrm{count}}=\Sigma (g_{0},\omega_{0},\phi _{0},\Phi
_{0})+O(\vartheta^{2})\;,  \label{stab}
\end{equation}
with the bare fields, sources and parameters defined as
\begin{align}
A_{0\mu }^{a} &=Z_{A}^{1/2}A_{\mu }^{a}\;, & K _{0\mu}^{a}&~=~Z_{K }K _{\mu }^{a}\;,  &  g_{0}~&=~Z_{g}g\;, \nonumber \\
c_{0}^{a} &=Z_{c}^{1/2}c^{a}\;,           & L_{0}^{a}~&=~Z_{L}L^{a}\;, & \omega_{0}&~=~Z_{\omega }\omega \;.  \nonumber \\
\overline{c}_{0}^{a} &=Z_{\overline{c}}^{1/2}\overline{c}^{a}\;, & K_{0}~&=~Z_{K}K\;,  &    \nonumber \\
b_{0}^{a} &=Z_{b}^{1/2}b^{a}\;,  &  \eta_{0}~&=~Z_{\eta }\eta\;, \nonumber\\
&               &             K_{0 \mu \nu}~&=~Z_{K_{\mu\nu}}K_{\mu\nu}\;, \nonumber\\
&           &   \eta_{0 \mu \nu}~&=~Z_{\eta_{\mu \nu} }\eta_{\mu \nu}\;,
\end{align}
and $\vartheta$ the infinitesimal perturbation parameter. Notice that for consistency, we should find that
$Z_{K_{\mu\nu}}=Z_K$, $Z_{\eta_{\mu\nu}}=Z_\eta$.

The parameters $a_0 $, $a_{1}$, $a_{6}$, $a_7$, $a_{8}$, $a_9$ turn
out to be related to the renormalization of the gauge coupling
constant $g$, of the fields $A_{\mu }^{a}$, $c^{a}$ and of the sources $K$, $K_{\mu\nu}$ ,
$\omega$ according to
\begin{eqnarray}
Z_{g} &=&1-\vartheta \frac{a_0}{2}\;,  \nonumber \\
Z_{A}^{1/2} &=&1+\vartheta \left( \frac{a_0}{2}+a_{1}\right) \;,  \nonumber \\
Z_{K} &=&1+\vartheta \left( a_{6}- a_0 \right)\;,   \nonumber \\
Z_{\omega } &=&1+\vartheta \left( 2a_0 -2a_{6} - a_{8}\right) \;.
\end{eqnarray}
Concerning the other fields and the sources $K _{\mu }^{a}$,
$L^{a}$, it can be verified that they renormalize as
\begin{eqnarray}
Z_{\overline{c}}^{1/2} &=& Z_{c}^{1/2} = Z_A^{-1/2} Z_g = 1-\vartheta \frac{a_{1}}{2}\;, \nonumber \\
Z_{b}&=&Z_{A}^{-1}\;, \nonumber\\
Z_{K }&=&Z_{c}^{1/2}\;,  \nonumber\\
Z_{L} &=&Z_{A}^{1/2}\;, \nonumber\\
Z_{\eta} &=& Z_{\eta_{\mu\nu}}  =  Z_K Z_A^{1/2}Z_c^{1/2} =  1 + \vartheta \left( a_6 - \frac{a_0}{ 2}+\frac{a_1}{ 2}\right)\;,\nonumber\\
Z_{ K_{\mu\nu} } &=& Z_{K}\;.
\end{eqnarray}
This completes the proof of the multiplicative renormalizability of
the LCO $\mathcal{O}_{\mu\nu}$ in the Landau gauge: the action \eqref{sym} is renormalizable, where the $Z$-factor of $K_{\mu\nu}$ is equal to the $Z$-factor of $K$, as required.

\subsection{Algebraic proof of the renormalizability of the local operator  $\mathcal{O}_{\mu\nu}$ in combination with the LCO $A^2$}
\subsubsection{The action}
The current problem is that the operator $A_\mu^2$ cannot be studied with the action \eqref{sym}. Indeed, if we set $K_{\mu\nu} = K \delta_{\mu\nu}$ and consequently $\eta_{\mu\nu} = \eta \delta_{\mu\nu}$, then the action $S_K = 0$. As mentioned in the Introduction, from a physical point of view, we also need $A_\mu^2$ as operator, therefore we consider the following action:
\begin{eqnarray}
\Sigma' &=&S_{YM}+S_{GF}+ S_K + S_{ext} + S_A  \label{sym2}\;,
\end{eqnarray}
with
\begin{eqnarray}  \label{renbis2}
S_{A} & = & s\int\d^{d}x\left(  \frac{1}{2}\lambda A_\mu^a A_\mu^a -
\frac{1}{2}\zeta \lambda J \right) \nonumber\\
&=& \int\d^{d}x\left( \frac{1}{2}J A_\mu^a A_\mu^a + \lambda A \p c
- \frac{1}{2}\zeta  J^2 \right) \;,
\end{eqnarray}
In this fashion, we coupled $A^2$ again to the action, with a doublet $(\lambda, J )$:
\begin{eqnarray}
s\lambda&=& J \;,\nonumber\\
s J &=&0\;.
\end{eqnarray}
The action \eqref{sym2} allows us to study the LCOs $A_\mu^2$ and $\mathcal{O}_{\mu\nu}$. Clearly, these 2 operators correspond to the decomposition of $A_\mu A_\nu$ into its trace and traceless components.

\subsubsection{Ward identities and the counterterm}
The action $\Sigma'$ obeys the same Ward identities as the action $\Sigma$ (see eq.~\eqref{ca}), only the Slavnov-Taylor identity is slightly modified:
\begin{equation}
\mathcal{S}(\Sigma' )=\int\d^d x\left( \frac{\delta \Sigma' }{\delta
A_{\mu
}^{a}}\frac{\delta \Sigma' }{\delta K ^a _{\mu }}+\frac{\delta \Sigma' }{\delta c^{a}}\frac{\delta \Sigma' }{\delta L^{a}}+b^{a}\frac{\delta \Sigma' }{ \delta \overline{c}^{a}}+K\frac{\delta \Sigma' }{\delta \eta}  +K_{\mu\nu} \frac{\delta \Sigma' }{\delta \eta_{\mu\nu}} +   J  \frac{\delta \Sigma' }{\delta \lambda}  \right)  =0\;. \label{sti}
\end{equation}
and we also have an extra identity,
\begin{eqnarray}
\int \frac{\delta \Sigma' }{\delta \lambda} + \int c \frac{\delta \Sigma' }{\delta b} &=& 0\;.
\end{eqnarray}
Therefore the counterterm obeys the following linearized Slavnov-Taylor identity
\begin{equation}
\mathcal{B}_{\Sigma }\Sigma ^{\prime \mathrm{count}}=0\;,  \label{b1}
\end{equation}
where $\mathcal{B}_{\Sigma }$
\begin{eqnarray}
&&\mathcal{B}_{\Sigma }=\int\d^d x\left( \frac{\delta \Sigma
}{\delta
A_{\mu }^{a}}\frac{\delta }{\delta K^a _{\mu }}+\frac{\delta \Sigma }{%
\delta K^a_{\mu }}\frac{\delta }{\delta A_{\mu }^{a}}+\frac{\delta
\Sigma }{\delta c^{a}}\frac{\delta }{\delta L^{a}}+\frac{\delta \Sigma }{%
\delta L^{a}}\frac{\delta }{\delta c^{a}}+b^{a}\frac{\delta }{\delta
\overline{c}^{a}}+K_{\mu\nu}\frac{\delta }{\delta \eta_{\mu\nu} } +
 J  \frac{\delta }{\delta \lambda }\right) \;,   \label{lb}
\end{eqnarray}
and it is also restricted by
\begin{eqnarray}
\int \frac{\delta \Sigma^{\prime count} }{\delta \lambda} &=& 0\;.
\end{eqnarray}
After imposing all the Ward
identities, we find for the counterterm
\begin{eqnarray}\label{count}
\Sigma ^{\prime \mathrm{count}} &=& \Sigma ^{ \mathrm{count}} + a_1
\int\d^d x \left( \frac{1}{2} J  A^a_\mu A^a_{\mu}  \right)
+\int\d^d x a_{11}\frac{\zeta}{2}    J^2\;,
\end{eqnarray}
with $\Sigma ^{ \mathrm{count}}$ given in \eqref{finalcount}. Notice
that, due to the additional Ward identity
\eqref{extrawardconstraints}, no mixing occurs between $ J $ and
$K$. This is a powerful result. A priori, a mixing between $\mathcal{O}_{\mu\nu}$ and $A_\kappa^2 \delta_{\mu\nu}$ cannot be excluded.

Absorbing the counterterm \eqref{count} back into the original action
$\Sigma'$ gives the additional $Z$-factors (all the others are
the same as before)
\begin{eqnarray}
Z_{\zeta } &=&1+\vartheta \left( 2a_0 +2a_{1} - a_{11}\right)\;, \nonumber\\
Z_{ J } &=&Z_g Z_A^{-1/2} = 1-\vartheta \left( a_{1}+ a_0 \right)\;,   \nonumber \\
Z_{\lambda} &=& Z_{ J } Z_A^{1/2}Z_c^{1/2}\;.
\end{eqnarray}
Summarizing our result so far, we have seen that we had to introduce 2 (independent) sources to
discuss the renormalization of $A_\mu A_\nu$ and $A_\mu^2$, and this
by means of the action $\Sigma'$. At the end of the story, we have
identified the two independently renormalizable operators
$\mathcal{O}_{\mu\nu}$ and $A_\mu^2$. This also means that the vacuum divergences $\sim
 J^2$ and the renormalization factor $Z_J$ remains unchanged
compared with the cases already studied in
\cite{Dudal:2002pq,Verschelde:2001ia,Browne:2003uv}.

\section{Preliminaries}
The next step will be to calculate the effective potential. Notice that this can be done relatively close along the lines of \cite{Verschelde:2001ia,Browne:2003uv}. We depart from the action
$\Sigma'$ and set all the external sources equal to zero, except $K_{\mu\nu}$ and $J$.

\subsection{Hubbard-Stratonovich transformation}\label{HS}
In this section, we shall get rid of the unwanted quadratic source dependence by the introduction of 2 Hubbard-Stratonovich fields. We start by writing down the complete action $\Sigma'$, where we have set all the external sources equal to zero, except $K_{\mu\nu}$ and $J$,
\begin{eqnarray}
\Sigma' &=& \frac{1}{4}\int\d^dxF_{\mu \nu }^{a}F^a_{\mu \nu }  +\int\d^dx\left( b^{a}\partial _{\mu }A^a_{\mu }+ \overline{c}^{a}\partial _{\mu }D_{\mu }^{ab}c^{b}\right)   \nonumber\\
&&+ \int\d^dx\left( \frac{1}{2} K_{\mu\nu} A_{\mu }^{a}A^{a}_\nu  -
\frac{1}{2d}  K A_{\mu }^{a}A^{a}_\mu  - \frac{\omega}{2} K_{\mu\nu}
K_{\mu\nu} + \frac{\omega }{2d}K^{2}  \right) + \int\d^{d}x\left(
\frac{1}{2}J A_\mu^a A_\mu^a  - \frac{1}{2}\zeta J^2 \right)\;.
\end{eqnarray}
The energy functional can be written as
\begin{eqnarray}\label{enfunc}
\e^{- W(J,K_{\mu\nu})} &=& \int [\d A_\mu][\d c][\d\overline c][\d b] \e^{-\Sigma'}\;.
\end{eqnarray}
We now rewrite the action as
\begin{equation}\label{enfunc2bis}
\Sigma' = S_{YM} +  S_{GF}+ \int\d^dx \left( \frac{1}{2}
k_{0,\mu\nu} A_{0\ \mu }^{a}A^{a}_{0\ \nu}  - \frac{1}{2} \omega_0
k_{0,\mu\nu}^2  \right) + \int\d^{d}x\left( \frac{1}{2}J_0 A_{0\
\mu}^a A_{0\ \mu}^a  - \frac{1}{2}\zeta_0 J_0^2 \right)\;,
\end{equation}
where we have used the abbreviation
\begin{eqnarray}
k_{\mu\nu} &=& K_{\mu\nu} - \frac{\delta_{\mu\nu}}{d} K\;,
\end{eqnarray}
and where we have used the bare quantities. We can then rewrite the
action in terms of finite fields and sources. We recall that in $d =
4 -\epsilon$ dimensions, we have the following mass dimensions,
\begin{eqnarray}
\left[A_\mu\right] &=& \frac{d-2}{2} = 1- \frac{\epsilon}{2}\;, \nonumber\\
\left[g\right] &=& \frac{4-d}{2} = \frac{\epsilon}{2}\;, \nonumber\\
\left[J \right] &=& \left[K\right] = 2\;, \nonumber\\
\left[\zeta\right] &=& \left[\omega\right] = d-4 = \epsilon\;.
\end{eqnarray}
With this in mind, the action $\Sigma'$ can be written as,
\begin{eqnarray}
\Sigma' &=&  S_{YM} +  S_{GF} + \int\d^dx \left[ \frac{1}{2} Z_K Z_A
k_{\mu\nu} A_{\mu }^{a}A^{a}_{\nu}  -    \frac{1}{2}Z_\omega Z_K^2
\mu^{-\epsilon} \omega k_{\mu\nu}^2    \right] +
\int\d^{d}x\left[ \frac{1}{2}Z_J Z_A J A_{ \mu}^a A_{ \mu}^a  -
\frac{1}{2} Z_\zeta Z_J^2 \mu^{-\epsilon} \zeta J^2 \right]\;.
\end{eqnarray}
We shall now perform two Hubbard-Stratonovich transformations by multiplying expression \eqref{enfunc} with the following unities,
\begin{eqnarray}\label{unities}
1 &=& \int [\d \sigma] \e^{-\frac{1}{2 Z_\zeta Z_J^2 \zeta} \int\d^d x \left( \frac{\sigma}{g} + \frac{1}{2} \mu^{\epsilon/2} Z_J Z_A A^2 - \mu^{-\epsilon/2} Z_\zeta Z_J^2 \zeta J \right)^2 }\;, \nonumber\\
1 &=& \int [\d \phi_{\mu\nu}] \e^{- \frac{1}{2 Z_\omega Z_K^2
\omega} \int\d^d x \left( \frac{1}{g} \left(  \phi_{\mu\nu} - \frac{\delta_{\mu\nu} }{d} \phi \right)  +
\frac{1}{2} \mu^{\epsilon/2} Z_A Z_K A_\mu A_\nu -
\mu^{-\epsilon/2} Z_\omega Z_K^2 \ \omega\ k_{\mu\nu} \right)^2}\;,
\end{eqnarray}
where
we have introduced two new fields, $\sigma$ and $\phi_{\mu\nu}$. We used a specific (traceless) combination, $\phi_{\mu\nu} - \frac{\delta_{\mu\nu} }{d} \phi$ with $\phi=\phi_{\kappa\kappa}$, the reason wherefore shall become clear soon. Let us
define the following abbreviation $\varphi_{\mu\nu}$,
\begin{eqnarray}
\varphi_{\mu\nu} &=& \phi_{\mu\nu} - \frac{\delta_{\mu\nu} }{d} \phi\;,
\end{eqnarray}
which is traceless, just like $k_{\mu\nu}$. Now performing the Hubbard-Stratonovich transformations yields
\begin{eqnarray}\label{enfunc2}
\e^{- W(J,K_{\mu\nu})} &=& \int [\d A_\mu][\d c][\d\overline c][\d b] [\d
\sigma][\d \phi_{\mu\nu}] \exp\left[- \int\d^d x \left( \mathcal
L(A_\mu, \sigma, \phi_{\mu\nu})   - \mu^{-\epsilon/2}
\frac{\sigma}{g}  J  - \mu^{-\epsilon/2}
\frac{\varphi_{\mu\nu}}{g} k_{\mu\nu} \right) \right]\;,
\end{eqnarray}
where
\begin{eqnarray}\label{sigmaactie}
\int \d^dx\;\mathcal L(A_\mu, \sigma, \phi_{\mu\nu}) &=&  S_{YM} +  S_{GF} +\int \d^dx\left[\frac{1}{2Z_{\zeta} Z_J^2 \zeta} \frac{\sigma^2}{g^2} +  \frac{1}{2 Z_\zeta Z_J \zeta g}  Z_A \mu^{\epsilon/2} \sigma A^a_\mu A^a_\mu + \frac{1}{8 Z_\zeta  \zeta} \mu^{\epsilon} Z_A^2 (A^a_\mu A^a_\mu)(A^b_\nu A^b_\nu)\right. \nonumber\\
&& \left.+ \frac{1}{2Z_{\omega} Z_K^2\omega}
\frac{\varphi_{\mu\nu}^2}{g^2} +  \frac{1}{2 Z_\omega Z_K \omega g}
Z_A \mu^{\epsilon/2} \varphi_{\mu\nu} A_\mu A_\nu + \frac{1}{8
Z_\omega \omega} \mu^{\epsilon} Z_A^2 (A^a_\mu A^a_\nu)(A^b_\mu A^b_\nu)\right]\;.
\end{eqnarray}
Notice that by the Hubbard-Stratonovich transformation, we have
removed the terms $\sim \zeta J^2$ respectively $\sim \omega
k_{\mu\nu}^2$ and $\sim J A^2$ respectively $\sim k_{\mu\nu}  A_\mu
A_\nu$. The sources $J$ and $k_{\mu\nu}$ are now linearly coupled to
$\sigma/g$ and $\varphi_{\mu\nu}/g$ as one can observe in equation
\eqref{enfunc2}. Hence, the usual 1PI formalism applies, with the following identification,
\begin{eqnarray}
% \nonumber to remove numbering (before each equation)
  \braket{\sigma} &=& -\frac{g}{2}\braket{A_\mu^2}\;, \nonumber\\
  \braket{\varphi_{\mu\nu}} &=& -g\braket{\mathcal{O}_{\mu\nu}}\;,
\end{eqnarray}
which follows easily from acting with $\frac{\p}{\p J}$ and $\frac{\p}{\p K_{\mu\nu}}$ on the equivalent generating functionals $\e^{-\Sigma'}$ and $\e^{-W(J,K_{\mu\nu})}$ and setting $J=0$, $K_{\mu\nu}=0$.

Here we can also clearly appreciate the role of the traceless combination $\varphi_{\mu\nu}$ used in \eqref{unities}: it ensures that in the final action \eqref{sigmaactie} the traceless combination $\frac{1}{2}A_{\mu}A_\nu-\frac{1}{2}\frac{\delta_{\mu\nu}}{d}A_\kappa A_\kappa\equiv {\cal O}_{\mu\nu}$ appears, as
\begin{eqnarray}
\frac{1}{2}\varphi_{\mu\nu} A_{\mu\nu}&=&\frac{1}{2}\left(\phi_{\mu\nu}-\frac{\delta_{\mu\nu}\phi_\kappa^2}{d}\right)A_{\mu\nu}~=~\frac{1}{2}\phi_{\mu\nu}\left(A_{\mu}A_\nu-\frac{\delta_{\mu\nu}}{d}A_\kappa^2\right)~=~\frac{1}{2}\left(\phi_{\mu\nu}-\frac{\delta_{\mu\nu}}{d}\phi_\kappa^2\right)\left(A_{\mu}A_\nu-\frac{\delta_{\mu\nu}}{d}A_\kappa^2\right)~=~\varphi_{\mu\nu}\mathcal{O}_{\mu\nu}\;.
\end{eqnarray}

\subsection{Determination of the LCO parameters $\zeta$ and $\omega$}
To this point, we did not yet determine the two LCO parameters $\zeta$ and $\omega$. In this section we shall do so by deriving a differential equation for $\zeta$ and $\omega$, in an analogous way as in \cite{Verschelde:2001ia}. In order to do so, we shall adapt the notation a bit, and define $\delta \zeta$ and $\delta \omega$ as follows,
\begin{eqnarray}\label{notatie}
\zeta_0 J_0^2 &=& \mu ^{-\epsilon} \left( 1 + \frac{\delta \zeta}{\zeta} \right)\zeta J^2 = \mu ^{-\epsilon} Z_\zeta Z_J^2 \zeta J^2\;,   \nonumber\\
\omega_0 \ k_{0,\mu\nu}^2 &=& \mu ^{-\epsilon} \left( 1 +
\frac{\delta \omega}{\omega} \right)\omega \ k_{\mu\nu}^2 = \mu
^{-\epsilon}  Z_\omega Z_K^2 \omega \ k_{\mu\nu}^2\;.
\end{eqnarray}
We further define the following anomalous dimensions,
\begin{eqnarray}\label{andim}
\mu \frac{\p}{\p \mu} g^2 &=& \beta(g^2)\;, \nonumber\\
\mu \frac{\p}{\p \mu} \ln Z_J &=& \gamma_J(g^2) \quad \Rightarrow \quad \mu \frac{\p}{\p \mu} J = - \gamma_J(g^2)J\;, \nonumber\\
\mu \frac{\p}{\p \mu} \ln Z_{K_{\mu\nu}} &=& \gamma_{K}(g^2) \quad
\Rightarrow \quad \mu \frac{\p}{\p \mu} K_{\mu\nu} = -\gamma_K(g^2)
K_{\mu\nu}\;.
\end{eqnarray}
Let us first consider the case without $\mathcal{O}_{\mu\nu}$, i.e. we set $K_{\mu\nu}\equiv0$. To determine a differential equation for $\zeta$, we take the derivative of the first equation of \eqref{notatie} w.r.t.~$\mu$,
\begin{eqnarray}
- \epsilon (\zeta + \delta \zeta) + \left( \mu \frac{\p}{\p \mu}
\zeta + \mu \frac{\p}{\p \mu} (\delta \zeta) \right) - 2
\gamma_J(g^2) (\zeta + \delta \zeta) &=& 0\;.
\end{eqnarray}
As we can consider $\zeta$ as a function of $g^2$, we can rewrite the previous equation as
\begin{eqnarray}\label{diff1}
&& \mu \frac{\p}{\p \mu} \zeta =  \epsilon (\zeta + \delta \zeta) - \beta(g^2)  \frac{\p}{\p g^2} (\delta \zeta)  + 2 \gamma_J(g^2) (\zeta + \delta \zeta) \nonumber\\
&\Rightarrow& \beta(g^2)  \frac{\p}{\p g^2} \zeta (g^2) =  2
\gamma_J(g^2) \zeta (g^2) + f(g^2)\;,
\end{eqnarray}
with
\begin{eqnarray}\label{delta1}
f(g^2) &=& \epsilon  \delta \zeta - \beta(g^2)  \frac{\p}{\p g^2}
(\delta \zeta) + 2 \gamma_J(g^2) \delta \zeta
\end{eqnarray}
a finite function of $g^2$. This particular choice of $\zeta(g^2)$ is the unique one which ensures a linear renormalization group equation for the generating functional
$W(J)$ while keeping multiplicative renormalizability for $\zeta$; \eqref{diff1} is solved with a Laurent series in $g^2$,
 \begin{equation}\label{lco}
 \zeta(g^2)=\frac{\zeta_0}{g^2}+\zeta_1+\zeta_2 g^2+\mathcal{O}(g^4)
 \end{equation}
 by keeping in mind that the $\beta$-function starts at order $g^4$, while a typical anomalous dimension at order $g^2$. Explicit calculations show that $f(g^2)$ at order $g^0$. Notice also that \eqref{diff1} implies that $\beta(g^2)$ and $\gamma_J(g^2)$ have to be known to $(n+1)$ loops if $\zeta(g^2)$ is to be known at $n$ loops. We refer the reader to the literature for all details involved in the LCO procedure \cite{Verschelde:1995jj,Verschelde:2001ia,Dudal:2002pq}.

We recall here from \cite{Verschelde:2001ia,Browne:2003uv} that we know up to three loops,
\begin{eqnarray}
\delta \zeta &=& \frac{N^2 - 1}{16 \pi^2} \left[ -\frac{3}{\epsilon} + \left( \frac{35}{2} \frac{1}{\epsilon^2} -  \frac{139}{6} \frac{1}{\epsilon} \right)\left(  \frac{g^2 N}{16\pi^2} \right) +   \left(-\frac{665}{6} \frac{1}{\epsilon^3} + \frac{6629}{36} \frac{1}{\epsilon^2} - \left(  \frac{71551}{432} + \frac{231}{16} \zeta(3) \right)  \frac{1}{\epsilon} \right)\left(  \frac{g^2 N}{16\pi^2} \right)^2 \right]
\end{eqnarray}
and
\begin{eqnarray}
 Z_J &=& 1- \frac{35}{6} \frac{1}{\epsilon} \left( \frac{g^2 N }{16 \pi^2}\right)  + \left[ \frac{2765}{72}  \frac{1}{\epsilon^2}-\frac{449}{48}  \frac{1}{\epsilon}   \right] \left( \frac{g^2 N }{16 \pi^2}\right)^2+ \left[ -\frac{113365}{432}\frac{1}{\epsilon^3} +  \frac{41579}{576}  \frac{1}{\epsilon^2}+\left(-\frac{75607}{2592}-\frac{3}{16}\zeta(3)\right)\frac{1}{\epsilon}  \right] \left( \frac{g^2 N }{16 \pi^2}\right)^3\;,\nonumber\\
\end{eqnarray}
so that
\begin{eqnarray}\label{ress1}
\gamma_J(g^2) &=&  \frac{35}{6}\left(  \frac{g^2 N}{16\pi^2} \right) +  \frac{449}{24}\left(  \frac{g^2 N}{16\pi^2} \right)^2 +  \left( \frac{94363}{864} - \frac{9}{16} \zeta(3) \right)\left(  \frac{g^2 N}{16\pi^2} \right)^3\;.
\end{eqnarray}
We can now do the same for the second equation of \eqref{notatie}. Notice that from equation \eqref{andim} follows
\begin{eqnarray}
\mu  \frac{\p}{\p \mu}  K_{\mu\mu} &=& - \gamma_K(g^2)K_{\mu\mu}\;,
\end{eqnarray}
and therefore,
\begin{eqnarray}
\mu \frac{\p}{\p \mu} k_{\mu\nu} &= - \gamma_K(g^2)k_{\mu\nu} \;.
\end{eqnarray}
Hence, we can derive an analogous differential equation for $\omega$
\begin{eqnarray}\label{diff2}
\beta(g^2)  \frac{\p}{\p g^2} \omega (g^2) &= & 2 \gamma_K(g^2) \omega
(g^2) + h(g^2)\;,
\end{eqnarray}
as we can also make $\omega$ as a function of $g^2$, where now
\begin{eqnarray}\label{delta2}
h(g^2) &=& \epsilon  \delta \omega - \beta(g^2)  \frac{\p}{\p
g^2} (\delta \omega) + 2 \gamma_K(g^2) \delta \omega\;.
\end{eqnarray}
At the end, the generating functional $W(J,K_{\mu\nu})$ will obey the following renormalization group equation:
\begin{equation}\label{RG0}
    \left(\mu\frac{\p}{\p\mu}+\beta(g^2)\frac{\p}{\p g^2}-\gamma_J(g^2)\int \d^4x\;J\frac{\delta}{\delta J}-\gamma_K(g^2)\int \d^4x\;K_{\mu\nu}\frac{\delta}{\delta K_{\mu\nu}}\right)W(J,K_{\mu\nu})~=~0\;,
\end{equation}
which in turn ensures a linear renormalization group equation for the effective action $\Gamma(\sigma,\phi_{\mu\nu})$. There is no explicit reference anymore
to neither $\zeta$ nor $\omega$, as these LCO parameters obey their renormalization group running by construction.

\subsection{Loop calculations }
In this subsection we discuss the determination of the anomalous dimension of
the ${\cal O}_{\mu\nu}$ operator to three loops in the $\MSbar$ scheme, as well
as the divergence structure of its associated vacuum energy. These clearly
extend the analogous results for the original $\half A_\mu^{a\;2}$, which were
derived in \cite{Verschelde:2001ia,Browne:2003uv}. Whilst we follow the same procedures here as \cite{Browne:2003uv},
there are some novel features associated with treating an operator with free
Lorentz indices, especially since it is also traceless. First, we note that we
take the same general point of view as \cite{Browne:2003uv} and include $\Nf$ flavours of
massless quarks and for the moment also consider the operator in an arbitrary
linear covariant gauge with parameter $\alpha$. At the end we will return to
Yang-Mills and the Landau gauge $\alpha$~$=$~$0$. The reason for taking the
more general scenario is that ironically with several additional parameters
present the renormalization constants we will determine have more internal
consistency checks. This gives us greater confidence in the final explicit
expressions. Moreover, as the operator has free Lorentz indices, not only is
there no operator mixing with the operator $\half A_\mu^{a\;2}$, unlike that
operator there is equally no mixing with any Faddeev-Popov ghost dependent
operator. Therefore, the renormalization of the operator ${\cal O}_{\mu\nu}$ as defined in eq.\eqref{operator} is multiplicative. We
will use dimensional regularization in $d$~$=$~$4$~$-$~$2\bar{\epsilon}$
dimensions to exploit the power of the calculational machinery of the
{\sc Mincer} algorithm, \cite{3}, where $\bar{\epsilon}$~$=$~$\half\epsilon$ in
this article. The algorithm is encoded, \cite{4}, in the symbolic manipulation
language {\sc Form}, \cite{5}, and is the only viable means to achieve the goal
of the three loop renormalization constants and anomalous dimension in a
significantly reasonable amount of time. We also note that having the three
loop results available at this stage, aside from gaining confidence in the
correctness of the two loop results, means that the groundwork is actually laid
for the future extension of our effective potential to two loops.

First, to renormalize ${\cal O}_{\mu\nu}$ we insert it into a gluon $2$-point
function and nullify the momentum of one external gluon leg. This is in order
to comply with the conditions of the {\sc Mincer} algorithm which determines
three loop scalar massless $2$-point functions to the finite part in
dimensional regularization. Concerning the gluon external leg momentum
nullification, given that each triple gluon vertex carries a numerator
momentum, this procedure does not introduce any spurious infrared singularities
which could plague, say, a similar procedure in a scalar field theory. By
contrast, the renormalization of $\half A_\mu^{a\;2}$ proceeded by inserting
the extended BRST invariant operator in a ghost $2$-point function, \cite{6,7}.
This reduced the number of internal gluon lines, resulting in a significantly
fast evaluation, since {\sc Mincer} relies heavily on integration by parts. An
increase in internal momenta slows programme speed. Whilst we could insert
${\cal O}_{\mu\nu}$ into a ghost $2$-point function, with appropriate
momentum nullification, the three loop {\sc Mincer} algorithm could only
produce two loop renormalization constants for the operator renormalization
constant. We therefore have to choose the gluon $2$-point function for the
operator renormalization and accept slower run times. Either way, however, one
would still first have to give the {\sc Mincer} routine scalar integrals. For
the Green function we consider, $\langle A^a_\sigma(p)
{\cal O}_{\mu\nu}(-p) A^b_\rho(0) \rangle$, there are four free Lorentz
indices and therefore one first needs to construct a Lorentz tensor projector.
Ordinarily one would derive the most general tensor objects built from the
momentum $p^\mu$ and the metric tensor, $\delta^{\mu\nu}$, with four free indices
in such a way that the tensor is consistent with the tracelessness and symmetry
of the free indices of the operator. Clearly there will be more than one such
independent tensor. Given whatever choice of basis is made, the scalar
amplitude associated with each independent tensor can be deduced by inverting
the tensor basis. However, as we are not actually interested in the finite
parts of the Green function $\langle A^a_\sigma(p) {\cal O}_{\mu\nu}(-p)
A^b_\rho(0) \rangle$ but only its divergence structure we need only find one
projector. This should be chosen to give a non-trivial tree contribution to
ensure there is a counterterm available to absorb the infinity which defines
the operator renormalization. Moreover, it ought to be chosen so as not to
increase computer run times, especially for the three loop Feynman diagrams.
Given these considerations we have projected with the tensor
\begin{equation}
{\cal P}^{\cal O}_{\mu\nu|\sigma\rho}(p) ~=~ \delta_{\mu\sigma}
\delta_{\nu\rho} ~+~ \delta_{\mu\rho} \delta_{\nu\sigma} ~-~ 2 \delta_{\sigma\rho}
\frac{p_\mu p_\nu}{p^2}~.
\end{equation}
The Feynman diagrams contributing to $\langle A^a_\sigma(p)
{\cal O}_{\mu\nu}(-p) A^b_\rho(0) \rangle$ are generated by the {\sc Qgraf}
package, \cite{8}, and the output converted to {\sc Form} input notation by
appending colour, Lorentz and spinor indices as well as the internal momenta
consistent with the {\sc Mincer} topology definitions, \cite{5}. For this
calculation there are $2$ one loop, $42$ two loop and $1023$ three loop Feynman
diagrams to be calculated. We follow the algorithm of \cite{9} to extract the
renormalization constant associated with ${\cal O}_{\mu\nu}$ defined by
\begin{equation}
{\cal O}_{0 \; \mu \nu} ~=~ Z^{{\cal O}}
{\cal O}_{\mu\nu}
\end{equation}
where ${}_0$ denotes a bare object. This algorithm,
\cite{9}, in essence is such that one computes the Green functions for bare
parameters and then scales to renormalized quantities at the end via the usual
definitions. The overall divergence which remains after the external fields
have been renormalized is absorbed by the as yet undetermined renormalization
constant $Z^{{\cal O}}$. As this renormalization is multiplicative then this
is easy to implement automatically within a {\sc Form} programme. The result of
our computation is the $\MSbar$ expression for the anomalous dimension
\begin{eqnarray}\label{res1}
 \gamma_K(g^2) &=& 2 \left[\frac{29 }{12}\frac{g^2 N}{16 \pi^2 } + \frac{389}{48} \left( \frac{g^2 N}{16 \pi^2 } \right)^2 - \frac{2754 \zeta(3) - 196111}{5184} \left( \frac{g^2 N}{16 \pi^2 } \right)^3\right]
\end{eqnarray}
with corresponding renormalization factor
\begin{eqnarray}
\!\!\!\!\!\!\! Z_K &=& 1- \frac{29}{6} \frac{1}{\epsilon} \left( \frac{g^2 N }{16 \pi^2}\right)  + \left[ \frac{2117}{72}  \frac{1}{\epsilon^2}-\frac{389}{48}  \frac{1}{\epsilon}  \right] \left( \frac{g^2 N }{16 \pi^2}\right)^2+\left[-\frac{82563}{432}\frac{1}{\epsilon^3}+\frac{99627}{864}\frac{1}{\epsilon^2}+\frac{2754\zeta(3)-196111}{7776}\frac{1}{\epsilon}\right] \left( \frac{g^2 N }{16 \pi^2}\right)^3
\end{eqnarray}
However, we have given the full expression for arbitrary gauge parameter and
quark flavours in Appendix \ref{appA}. For that result and by implication
for (\ref{res1}), the anomalous dimension passes all the usual internal
consistency checks. Specifically the extracted $\MSbar$ renormalization
constant is a Laurent series in $\epsilon$. Therefore the double and triple
poles in $\epsilon$ are predicted by the simple poles from lower loop order due
to the renormalization group equation since we are dealing with a
renormalizable operator. Our renormalization constant successfully satisfied
this check. As an additional check on our computer code, such as the correct
{\sc Qgraf} generation of diagrams and mapping to {\sc Mincer} and {\sc Form}
syntax, we have rerun the complete code for the calculation again but instead
used the operator $\half A_\mu^{a\;2}$. We obtained the known three loop
anomalous dimension for the Landau gauge, \cite{6,7}. The arbitrary $\alpha$
expression in this instance is not actually a check for this purpose for
several reasons. First, the arbitrary $\alpha$ result which is available is
for the extension of the operator to the BRST invariant object and only for the
non-linear arbitrary covariant gauge known as the Curci-Ferrari gauge,
\cite{10}, rather than a {\em linear} covariant gauge. Aside from this, if one
merely considered $\half A_\mu^{a\;2}$ on its own in an abritrary linear
covariant gauge then there is a potential mixing to the Faddeev-Popov ghost
mass operator $\bar{c}^a c^a$ for $\alpha$~$\neq$~$0$. It is not necessary to
investigate this as we are only primarily interested in checking a working code
whose output satisfies several stringent checks already.

The second main three loop result which we have determined is the divergence
structure of the associated operator vacuum energy. As outlined in \cite{Browne:2003uv}
this can be deduced by considering a massless $2$-point function with the
operator present at each of the two external points and a non-zero momentum
flowing in one operator and out the other. The divergence structure of this
Green function reproduces the divergence which occurs in the vacuum graphs
composing the effective potential. In essence these vacuum diagrams involve the
constant current of the LCO formalism which gives massive propagators. By
formally differentiating with respect to these masses one accesses the part of
the vacuum diagrams which are divergent. This divergence can then be extracted
by formally setting the mass to zero in these cut open vacuum graphs to produce
the massless $2$-point function we compute, \cite{Browne:2003uv}. As the set-up we have
described involves massless $2$-point Feynman diagrams, it is naturally
accessible to the {\sc Mincer} algorithm. Moreover, one does not need to
nullify any external momenta. Therefore, we have generated the diagrams using
{\sc Qgraf}, which gives $1$ one loop, $7$ two loop and $127$ three loop
Feynman diagrams. However, as with the renormalization of ${\cal O}_{\mu\nu}$
we need to project with an appropriate Lorentz tensor to have scalar amplitudes,
since our Green function is in effect $\langle {\cal O}_{\mu\nu}(p)
{\cal O}_{\sigma\rho}(-p) \rangle$. Again only the divergence structure is
required and therefore we applied the projector
\begin{equation}
{\cal P}^K_{\mu\nu|\sigma\rho} ~=~ \frac{1}{2(d-1)(d+2)} \left[
\delta_{\mu\sigma} \delta_{\nu\rho} ~+~ \delta_{\mu\rho} \delta_{\nu\sigma} ~-~
\frac{2}{d} \delta_{\mu\nu} \delta_{\sigma\rho} \right]\;.
\end{equation}
Consequently we find the divergence is
\begin{eqnarray}\label{deltaomega}
\delta \omega &=& \frac{N^2 -1}{16 \pi^2} \left[-\frac{7}{12} \frac{1}{\epsilon} +  \left( \frac{203}{72} \frac{1}{\epsilon^2} - \frac{1345}{864} \frac{1}{\epsilon}\right)\frac{ g^2 N}{16 \pi^2} \right] +  \mathcal{O}(g^4)\;.
\end{eqnarray}
Again this is for Yang-Mills and we have devolved the full $\Nf$ expression for
the Landau gauge to Appendix \ref{appA}. Moreover, there we also give the arbitrary
$\Nf$ results for various quantities derived from these expressions whose
Yang-Mills versions appear subsequently in the main text hereafter.

\subsection{Solving the renormalization group equations for $\zeta(g^2)$ and $\omega(g^2)$} \label{sectionZfactoren}
For these calculations, we shall require the two loop $\beta$-function,
\begin{eqnarray}
\beta(g^2) &=& -\epsilon g^2 - 2\left(  \beta_0 g^4 + \beta_1 g^6 + O(g^8)  \right)\;,
\end{eqnarray}
with
\begin{eqnarray}
\beta_0 &=&  \frac{11}{3}  \left( \frac{N}{16 \pi^2} \right)\;,\qquad \beta_1 ~=~  \frac{34}{3}                \left( \frac{N}{16 \pi^2}\right)\;.
\end{eqnarray}
In fact, in \cite{Verschelde:2001ia,Browne:2003uv}, $\zeta$, $\delta \zeta$, and $Z_\zeta$ were already calculated, but for the benefit of the reader, we shall repeat the results here in a structured way, especially since we have used a slightly different notation than in \cite{Verschelde:2001ia}.

Firstly, combining $\gamma_J$ and $\delta \zeta$ into expression \eqref{delta1} yields,
\begin{eqnarray}\label{ress2}
f(g^2) &=& \frac{(N^2 -1)}{ 16 \pi^2}\left[ -3 - \frac{139}{3} \left( \frac{g^2 N}{16 \pi^2 } \right) - \left( \frac{71551}{144} + \frac{693}{16} \zeta(3) \right) \left( \frac{g^2 N}{16 \pi^2 } \right)^2 + O( (g^2)^3) \right]\;,
\end{eqnarray}
and analogously, we find $h(g^2)$ by combining $\gamma_K$ and $\delta \omega$ into expression \eqref{delta2},
\begin{eqnarray}\label{res2}
 h (g^2)&=& \frac{(N^2 -1)}{ 16 \pi^2}\left[ -\frac{7}{12} - \frac{1345}{432} \frac{g^2 N}{16 \pi^2 } - \frac{9881  + 8886\zeta(3)}{1152} \left( \frac{g^2 N}{16 \pi^2 } \right)^2 + O( (g^2)^4) \right]\;.
\end{eqnarray}
Secondly, solving the differential equation \eqref{diff1} and invoking expression \eqref{ress1} and \eqref{ress2} we can determine $\zeta$ to one loop order\footnote{In principle, we can go one loop further with the results known. However, we shall not need this next order, as we shall only determine the effective potential to one loop order. It will become clear that this is already a highly complicated task.},
\begin{eqnarray}
\zeta &=& \frac{N^2 -1}{16 \pi^2} \left[\frac{9}{13} \frac{16 \pi^2}{ g^2 N} + \frac{161}{52} \right]\;.
\end{eqnarray}
Likewise, solving the differential equation \eqref{diff2} and using \eqref{res1} and \eqref{res2}, we can determine $\omega$ to one loop order,
\begin{eqnarray}
\omega &=& \frac{N^2 -1}{16 \pi^2} \left[\frac{1}{4} \frac{16 \pi^2}{ g^2 N} + \frac{73}{1044} \right]\;.
\end{eqnarray}
Thirdly, from expression \eqref{notatie}, we can determine $Z_J^2 Z_\zeta$
\begin{eqnarray}\label{Z1}
Z_J^2 Z_\zeta &=& 1 + \frac{\delta \zeta}{ \zeta} = 1 - \frac{13}{3 \epsilon}   \left( \frac{g^2 N }{16 \pi^2}\right)+\mathcal{O}(g^4)\;,
\end{eqnarray}
and $Z_K^2 Z_\omega$,
\begin{eqnarray}\label{Z2}
Z_K^2 Z_\omega &=& 1 + \frac{\delta \omega}{ \omega} = 1 - \frac{7}{3 \epsilon}   \left( \frac{g^2 N }{16 \pi^2}\right)+\mathcal{O}(g^4)\;.
\end{eqnarray}

\section{The effective potential at zero temperature}
\subsection{Computation of the effective potential}
We are now ready to calculate the one loop effective potential. Firstly, from the results in section \ref{sectionZfactoren} we find that the tree level mass associated to $A^a_\mu A^a_\mu$ is
given by
\begin{eqnarray}
m^2 &=& \frac{13}{9} \frac{N}{ N^2 - 1 } \sigma g = \sigma' g\;,
\end{eqnarray}
where we have defined
\begin{eqnarray}
 \sigma' &=& \sigma \frac{13}{9} \frac{N}{ N^2 - 1 } \;.
\end{eqnarray}
Analogously, the tree level mass matrix associated with $A_\mu
A_\nu$ is given by,
\begin{eqnarray}
M_{\mu\nu} &=&  4 \frac{ N}{N^2 - 1} \varphi_{\mu\nu}  g =
\varphi'_{\mu\nu} g\;,
\end{eqnarray}
where we have defined
\begin{eqnarray}
 \varphi_{\mu\nu} ' &=&   4 \frac{ N}{N^2 - 1} \varphi_{\mu\nu} \;.
\end{eqnarray}
To calculate the effective potential for $\sigma'$, $\varphi'$, we
can rely on the background formalism. At one loop, only the
integration over the gluon field $A_\mu$ gives a nontrivial contribution, and we find
\begin{eqnarray}\label{effpot}
\e^{ - \Gamma^{(1)}(\sigma', \varphi_{\mu\nu}' )} &=& \int [\d A] \exp - \Biggl\{ \int \d^4 x \left[\frac{1}{2} A_{\mu}^a \delta^{ab} \left(  \delta_{\mu\nu} \p^2  + \left(1-\frac1\xi\right)\p_\mu \p_\nu + m^2 + M_{\mu\nu}  \right) A_\nu^b \right] \nonumber\\
&& \hspace{2cm}+ \int \d^4 x \left( \frac{1}{2Z_{\zeta} Z_J^2 \zeta} \frac{\sigma^2}{g^2}+ \frac{1}{2Z_{\omega} Z_K^2\omega} \frac{\varphi_{\mu\nu}^2}{g^2} \right) \Biggr\}
\nonumber\\
&=& \exp-\left(\frac{N^2 -1}{2} \tr \ln \mathcal Q_{\mu\nu} \right)  \exp -\left( \frac{1}{2Z_{\zeta} Z_J^2 \zeta} \frac{\sigma^2}{g^2}+ \frac{1}{2Z_{\omega} Z_K^2\omega} \frac{\varphi_{\mu\nu}^2}{g^2} \right)\;,
\end{eqnarray}
where $\tr \ln \mathcal  Q_{\mu\nu}$ is defined as,
\begin{eqnarray}\label{start}
\tr \ln \mathcal Q_{\mu\nu} &=& \tr\ln \left( -\p^2\delta_{\mu\nu} + \left(1-\frac1\xi\right)\p_\mu
\p_\nu + \delta_{\mu\nu}m^2 + M_{\mu\nu}\right)\;,
\end{eqnarray}
with $M_{\mu\nu}$ the traceless matrix which describes the asymmetry in $d$ dimensions.

Let us first start with the calculation of \eqref{start}. We can parameterize the traceless matrix $M_{\mu\nu}$ as follows,
\begin{equation}\label{param}
M_{\mu\nu} = A \begin{pmatrix} 1 &&& \\ &-\frac{1}{d-1}&& \\
&&\ddots&
\\ &&&-\frac{1}{d-1} \end{pmatrix}\;,
\end{equation}
as we can assume that there is still spatial symmetry when separating the temporal component. We can now rewrite expression \eqref{start},
\begin{eqnarray}
\tr \ln \mathcal Q_{\mu\nu} &=& \tr\ln\left( -\p^2\delta_{\mu\nu} + \left(1-\frac1\xi\right)\p_\mu
\p_\nu + \delta_{\mu\nu}m^2\right) + \tr\ln\left(\delta_{\mu\nu} +
\frac1{-\p^2+m^2} \left(\delta_{\mu\lambda}+(1-\xi)\frac{\p_\mu
\p_\lambda}{-\p^2+\xi m^2}\right) M_{\lambda\nu}\right)\;,
\end{eqnarray}
which becomes in the Landau gauge limit $\xi\rightarrow0$,
\begin{eqnarray}
\tr \ln  \mathcal Q_{\mu\nu} &=& (d-1)\tr\ln(-\p^2 + m^2) + \tr\ln ( -\p^2) + \tr\ln\left(\delta_{\mu\nu}+ \frac1{-\p^2+m^2} \left(\delta_{\mu\lambda}-\frac{\p_\mu \p_\lambda}{\p^2}\right) M_{\lambda\nu}\right)\;.
\end{eqnarray}
The matrix of the last logarithm has the following eigenvalues:  $1 -\frac1{-\p^2+m^2}\frac {A}{d-1}$ with multiplicity $(d-2)$, $1 + \frac
A{-\p^2+m^2}\left(1-\frac{d}{d-1}\frac{\p_0^2}{\p^2}\right)$ and $1$. Therefore, we obtain
\begin{eqnarray}
\tr \ln  \mathcal Q_{\mu\nu}&=&(d-1)\tr\ln(-\p^2 + m^2)+\tr\ln (-\p^2)+(d-2)\tr\ln\left(1-\frac1{-\p^2+m^2}\frac{A}{d-1}\right) \nonumber\\
&&+\tr\ln\left(1+\frac{A}{^-\p^2+m^2}\left(1-\frac{d}{d-1}\frac{\p_0^2}{\p^2}\right)\right) \nonumber\\
&=&  \underbrace{\tr\ln (-\p^2)}_{\mathrm{I}} + \underbrace{(d-2) \tr\ln\left(-\p^2+m^2-\frac {A}{d-1}\right)}_{\mathrm{II}}  + \underbrace{\tr\ln\left(-\p^2+m^2 +A\left(1-\frac{d}{d-1}\frac{\p_0^2}{\p^2}\right)\right)}_{\mathrm{III}}  \;.
\end{eqnarray}
Firstly, in dimensional regularization, we know that\footnote{An overall factor $(VT)$ will always be omitted in all the calculations. }
\begin{eqnarray}\label{logint}
\mathrm{I} = \tr\ln (-\p^2) &=& \int \frac{\d^d k}{(2\pi)^d} \ln k^2
= 0\;.
\end{eqnarray}
Secondly, part II is a readily evaluated as
\begin{eqnarray}
\mathrm{II} &=&(d-2)\tr\ln\left(-\p^2 +m^2-\frac A{d-1}\right)\nonumber\\
 &=& \frac1{(4\pi)^2} \left(m^2-\frac A3\right)^2\left(-\frac{2}{\epsilon} - \frac{1}{2} + \ln\frac{m^2-\frac A3}{\overline{\mu}^2} \right) +\frac{4A}{9(4\pi)^2} \left(m^2-\frac A3\right) + \mathcal
O(\epsilon)\;,
\end{eqnarray}
where we have worked in the $\MSbar$ scheme. Notice that, for II to be real valued, we have the following constraint:
\begin{eqnarray}\label{voorwaarde1}
m^2-\frac A3 &\geq& 0\;.
\end{eqnarray}
Finally, part III requires much more effort due to the presence of the temporal derivatives. In the appendix, we have worked out this calculation in great detail. In \eqref{partIII}, we have ultimately found,
\begin{eqnarray}
\mathrm{III} &=&
-\frac{1}{18(4\pi)^2} \left[5 A^2 + 12 A m^2 + 9 m^4 \right]\left[ \frac{2}{\epsilon} - \ln\left( \frac{A}{A-3m^2}\right) - \ln\left( \frac{m^2- A/3}{\overline{\mu}^2}\right) \right]\nonumber\\
&&  + \frac{1}{108(4\pi)^2} \left( -7 A^2 + 15 A m^2 + 27 m^4 + 27 \frac{m^6}{A}\right)  +\frac{ (m^2+A)^{5/2}}{15\pi^2\sqrt{m^2-\frac{A}{3}}}{}_2F_1\left(4,\frac{1}{2};\frac{7}{2};\frac{m^2+A}{m^2-\frac{A}{3}}\right)\;,
\end{eqnarray}
with ${}_2F_1$ a hypergeometric function. As one can find in the appendix, this part is real valued if
\begin{eqnarray}\label{voorwaarde2}
m^2+ A &\geq& 0 \;,
\end{eqnarray}
which is a second constraint. Taking parts II and III together yields,
\begin{eqnarray}\label{detfinal}
 \tr \ln Q_{\mu\nu} &=&\frac{1}{18(4\pi)^2} \left[  -\frac{2}{\epsilon}  +  \ln\left( \frac{m^2- A/3}{\overline{\mu}^2}\right)\right] \left[ 7A^2 + 27 m^4  \right] + \frac{1}{108(4\pi)^2} \left[ -29 A^2 + 99 A m^2 - 27 m^4 + 27 \frac{m^6}{A} \right]  \nonumber\\
&& +\frac{1}{18(4\pi)^2}\left[5 A^2 + 12 A m^2 + 9m^4 \right]\ln\left( \frac{A}{A-3m^2}\right)   +\frac{ (m^2+A)^{5/2}}{15\pi^2\sqrt{m^2-\frac{A}{3}}}{}_2F_1\left(4,\frac{1}{2};\frac{7}{2};\frac{m^2+A}{m^2-\frac{A}{3}}\right)\;.
\end{eqnarray}
The remaining hypergeometric function can be expressed in terms of elementary functions, since
\begin{eqnarray}\label{elem}
{}_2F_1\left(4,\frac{1}{2};\frac{7}{2};z\right)&=&\frac{5}{128}\frac{1}{z-1}\frac{1}{z^{5/2}}
\left(\sqrt{z}\left(-15z^2+4z+3\right)+\frac{3}{2}\left(5z^3-3z^2-z-1\right)\left(\ln(1+\sqrt{z})-\ln(1-\sqrt{z})\right)\right)\;.
\end{eqnarray}
In our case, we have $z\equiv\nu$ and we have 3 cases. Let us first
consider $z>1$, or $A>0$. In this case, the $\ln\left(
\frac{A}{A-3m^2}\right)$ picks up an imaginary part, leading to
\begin{eqnarray}
\mathrm{Im}\left[\frac{1}{18(4\pi)^2}\left[5 A^2 + 12 A m^2 + 9 m^4
\right]\ln\left(
\frac{A}{A-3m^2}\right)\right]&=&\frac{1}{288\pi}\left(5 A^2 + 12 A
m^2 + 9 m^4 \right)\;.
\end{eqnarray}
Then, keeping the constraints \eqref{voorwaarde1},
\eqref{voorwaarde2} in mind and using \eqref{elem}, we also find
\begin{eqnarray}
 \mathrm{Im}\left[\frac{
(m^2+A)^{5/2}}{15\pi^2\sqrt{m^2-\frac{A}{3}}}{}_2F_1\left(4,\frac{1}{2};\frac{7}{2};\frac{m^2+A}{m^2-\frac{A}{3}}\right)\right]&=&-\frac{1}{288\pi}\left(5  A^2 + 12 A m^2 + 9m^4 \right)\;.
\end{eqnarray}
This is a first consistency check, as the imaginary parts
neatly cancel.

Secondly, for $z<1$, or $A<0$, all constituent
functions in the $\tr \ln$ are real to start with.

A third interesting test is the $z\to 1$ case, i.e. the $A\to 0$
limit, which should give back the known result for
$(d-1)\mathrm{tr}\ln(-\p^2+m^2)$. All the terms in $1/A$ nicely
cancel, and we recover the correct value,
\begin{eqnarray}
 \tr \ln Q_{\mu\nu} &=& \frac{m^4}{16\pi^2}\left(-\frac{5}{4}-\frac{3}{\epsilon}+\frac{3}{2}\ln\frac{m^2}{\overline{\mu}^2}\right)\;.
\end{eqnarray}

The second part of expression \eqref{effpot} is, to the orders in $g^2$ and $\epsilon$ we need, readily evaluated as
\begin{eqnarray}
\frac{1}{2Z_{\zeta} Z_J^2 \zeta} \frac{\sigma^2}{g^2} &=& \frac{9}{26} \frac{N^2 - 1}{N} (\sigma')^2 \left[ 1 + \frac{13}{3 \epsilon} \left( \frac{g^2 N}{16 \pi^2} \right) - \frac{161}{36}  \left( \frac{g^2 N}{16 \pi^2} \right) \right] \nonumber\\
&=& \frac{9}{26} \frac{N^2 - 1}{N} \frac{m^4}{g^2} \left[ 1 + \frac{13}{3 \epsilon} \left( \frac{ N g^2}{16 \pi^2} \right) - \frac{161}{36}  \left( \frac{ g^2N}{16 \pi^2} \right) \right]\;, \nonumber\\
\frac{1}{2Z_{\omega} Z_K^2\omega} \frac{\varphi_{\mu\nu}^2}{g^2}  &=&\frac{1}{8} \frac{N^2 - 1}{N} (\varphi_{\mu\nu}')^2 \left[ 1 + \frac{7}{3 \epsilon} \left( \frac{g^2 N}{16 \pi^2} \right) - \frac{73}{261}  \left( \frac{g^2 N}{16 \pi^2} \right) \right] \nonumber\\
& =&\frac{1}{8} \frac{N^2 - 1}{N} \frac{A^2}{g^2} \left( \frac{4}{3} + \frac{\epsilon}{9} \right) \left[ 1 + \frac{7}{3 \epsilon} \left( \frac{g^2 N}{16 \pi^2} \right) - \frac{73}{261}  \left( \frac{g^2 N}{16 \pi^2} \right) \right]\;,
\end{eqnarray}
where we have used the results of the previous subsection \ref{sectionZfactoren}, in particular \eqref{Z1} and \eqref{Z2}.

In summary, taking all the results together, we nicely find that all the divergences cancel in the effective potential,
which, after some algebra and simplification leads to the final result,
\begin{eqnarray}\label{potentiaal}
 V^{(1)}(m, A)  &=& \frac{N^2 - 1}{2 (4\pi)^2} \Biggl\{ \frac{1}{18}  \ln\left( \frac{m^2- A/3}{\overline{\mu}^2}\right) \left[ 7A^2 + 27 m^4  \right] +  \left[ -\frac{155}{522} A^2 + \frac{11}{12} A m^2 - \frac{87}{26} m^4 + \frac{1}{4} \frac{m^6}{A} \right] \nonumber\\
&& +\frac{1}{18}\left[5 A^2 + 12 A m^2 + 9m^4 \right] \left[ \ln\left( \frac{A}{A-3m^2}\right) +\ln\left(1+\sqrt{ \frac{m^2+A}{m^2-\frac{A}{3}}}\right)-\ln\left(1-\sqrt{ \frac{m^2+A}{m^2-\frac{A}{3}}}\right) \right] \nonumber\\
  &&- \frac{\left(m^2 - \frac{A}{3}\right)}{12 A} (6 A^2 + 11 A m^2 + 3 m^4) \sqrt{ \frac{m^2+A}{m^2-\frac{A}{3}}}  + \frac{9}{13} \frac{(4\pi)^2}{g^2 N} m^4 + \frac{1}{3} \frac{(4\pi)^2}{g^2 N} A^2  \Biggr\}\;.
\end{eqnarray}

\subsection{A check: a handmade calculation of $\delta\omega$ at lowest order}
Having determined the logarithmic determinant in the previous subsection, we are now also in a
position to provide a handmade calculation of  the counterterm $\delta\omega$, to have a
check on the automated output leading to \eqref{deltaomega}. First, we single out the one loop divergence in $W(J=0,K_{\mu\nu})$,
as defined in \eqref{enfunc}. We can identify
$k_{\mu\nu}\equiv M_{\mu\nu}$, meaning that it is actually given by
the $\frac{1}{\epsilon}$ part of
\begin{equation}
\frac{N^2-1}{2}\tr\ln Q_{\mu\nu}\;.
\end{equation}
Using the conventions \eqref{enfunc} and \eqref{enfunc2bis}, this
divergence must be canceled by the counterterm contribution
\begin{equation}
-\frac{1}{2}\delta\omega
k_{\mu\nu}k_{\mu\nu}=-\frac{1}{2}\delta\omega A^2 \frac{d}{d-1}\;,
\end{equation}
with the parametrization \eqref{param}. Employing \eqref{detfinal}, we
find
\begin{equation}
\frac{N^2-1}{2}\tr\ln
Q_{\mu\nu}=-\frac{7}{9}\frac{N^2-1}{16\pi^2}\frac{A^2}{2\epsilon}+\mathrm{finite}\;,
\end{equation}
hence, at lowest order
\begin{equation}
\delta\omega
=-\frac{7}{9}\frac{d-1}{d}\frac{N^2-1}{16\pi^2}\frac{1}{\epsilon}=-\frac{7}{12}\frac{N^2-1}{16\pi^2}\frac{1}{\epsilon}\;,
\end{equation}
a value consistent with \eqref{deltaomega}.

Let us also include a check using the renormalization group equation. In particular,
we have for the complete one loop functional\footnote{We did not write down the counterterm contribution here.} $W^{(1)}(J,K_{\mu\nu})$,
\begin{equation}\label{RG1}
    W^{(1)}(J,K_{\mu\nu})=-\frac{1}{2}\frac{\omega_0}{g^2}k_{\mu\nu}^2-\frac{1}{2}\omega_1
    k_{\mu\nu}^2 -\frac{1}{2}\frac{\zeta_0}{g^2}J^2-\frac{1}{2}\zeta_1
    J^2+\frac{1}{2}(N^2-1)\tr\ln Q_{\mu\nu}\;,
\end{equation}
or
\begin{equation}\label{RG2}
    W^{(1)}(J,K_{\mu\nu})=-\frac{1}{2}\frac{\omega_0}{g^2}\frac{4A^2}{3}-\frac{1}{2}\omega_1
    \frac{3A^2}{4} +\frac{1}{36}\frac{N^2-1}{16\pi^2}\ln
    \frac{J-A/3}{\omu^2}(7A^2)+\textrm{terms in $J$, higher order terms }\;.
\end{equation}
Focusing on the $A$-sector, then the application of the renormalization group equation \eqref{RG0} leads to
\begin{equation}\label{RG2}
    \mu\frac{\d}{\d\mu}W^{(1)}(J,K_{\mu\nu})=-\frac{1}{2}\frac{4}{3}A^2\frac{\omega_0}{g^4}2\beta_0
    g^4+\frac{1}{2}\frac{4}{3}A^2\frac{\omega_0}{g^2}2\gamma_{K,0}+\frac{7A^2}{36}\frac{N^2-1}{16\pi^2}(-2)+\mathcal{O}(g^4)
    =0 + \mathcal{O}(g^4)\;,
    \end{equation}
after filling in the numbers for $\omega_0$, $\beta_0$ and
$\gamma_{K,0}$, since the anomalous dimension of $A$ is determined
by $-\gamma_K(g^2)$, which follows directly from the definition of $A$. We have thus explicitly verified at lowest order that the
definition of the LCO parameters $\omega$ and $\zeta$ gives a generating functional $W(J,K_{\mu\nu})$ consistent with the renormalization group equation.

\subsection{Minimum of the potential}
We shall now determine the minimum of the potential \eqref{potentiaal}, by solving the
associated gap equations
\begin{eqnarray}\label{condities}
\frac{\p V^{(1)}}{\p m^2} &=& 0\;, \nonumber\\
\frac{\p V^{(1)}}{\p A} &=& 0\;,
\end{eqnarray}
and checking which potential extrema is a minimum.

Before doing this, we can already check whether for small $A$, the
minimum $m^2 = (2.03\lms)^2$ from
\cite{Verschelde:2001ia,Browne:2003uv} is a stable one. A Taylor
expansion around $A=0$
yields
\begin{eqnarray}\label{potentiaal}
 V^{(1)}(m, A)  &=& \underbrace{\frac{N^2-1}{32\pi^2}\left\{-\frac{113 m^4}{26}+\frac{144 m^4 \pi ^2}{13 g^2 N}+\frac{3}{2} m^4 \ln \left[\frac{m^2}{\overline \mu^2}\right]\right\}}_{V^{(1)}(m, A=0)} + \frac{N^2-1}{32\pi^2}\left\{\left(\frac{85}{3132}+\frac{16 \pi ^2}{3 g^2 N}+\frac{7}{18} \ln \left[\frac{m^2}{\overline \mu^2}\right]\right)
 A^2\right\}+\mathcal{O}(A^3)\;.\nonumber\\
\end{eqnarray}
The first part is exactly the first order potential found in
\cite{Verschelde:2001ia,Browne:2003uv}, and the second part is a
positive correction\footnote{We recall that we set $\omu^2=m^2$ in
\cite{Verschelde:2001ia,Browne:2003uv} based on the renormalization
group.} in $A^2$. Moreover, one can also check that $\frac{\p^2
V^{(1)}}{\p m^2 \p A}=\frac{\p^2 V^{(1)}}{\p A \p m^2}=0$ at
\begin{eqnarray}\label{min1}
A=0 \;,\quad m^2 \approx 4.12 \ \lms^2
\end{eqnarray}
while it is already known from
\cite{Verschelde:2001ia,Browne:2003uv} that $\frac{\p^2 V^{(1)}}{\p
(m^2)^2}>0$ at \eqref{min1}. This means that $\eqref{min1}$ is a
stable (local) minimum, using the Hessian determinant. This is a
nice and nontrivial verification of the results of
\cite{Verschelde:2001ia,Browne:2003uv}, while
establishing Lorentz invariance (as $A=0$), a \emph{conditio sine qua non} at $T=0$.

Let us now look for other minima. The gap equations \eqref{condities} yield
\begin{eqnarray}
0&=&\frac{7 A}{12}-\frac{74 m^2}{13}+\frac{3 m^4}{4 A}+\sqrt{\frac{A+m^2}{3 m^2-A}} \left( \frac{5 A}{4\sqrt{3}}-\sqrt{3} m^2 -\frac{3 \sqrt{3} m^4 }{4 A}\right)+\frac{288 m^2 \pi ^2}{13 g^2 N} \nonumber\\
&& + \left(\frac{2}{3} A  + m^2 \right) \left(  \ln\left[\frac{A}{A-3 m^2}\right] - \ln\left[1-\sqrt{\frac{A+m^2}{-\frac{A}{3}+m^2}}\right]+  \ln\left[1+\sqrt{\frac{A+m^2}{-\frac{A}{3}+m^2}}\right] \right)+3 m^2 \ln\left[\frac{-\frac{A}{3}+m^2}{\overline \mu^2}\right]\;, \nonumber\\
0&=&-\frac{107 A}{522}+\frac{5 m^2}{4}+\frac{m^4}{2 A}-\frac{m^6}{4 A^2}+ \sqrt{\frac{A+m^2}{3 m^2-A}} \left( \frac{A}{\sqrt{3}}-\frac{29 m^2 }{12 \sqrt{3}}-\frac{2 m^4 }{\sqrt{3} A}+\frac{\sqrt{3} m^6}{4 A^2} \right)+\frac{32 A \pi ^2}{3 g^2 N} \nonumber\\
&&+ \left(\frac{5}{9} A +\frac{2}{3} m^2\right) \left(\ln\left[\frac{A}{A-3 m^2}\right] - \ln\left[1-\sqrt{\frac{A+m^2}{-\frac{A}{3}+m^2}}\right]+ \ln\left[1+\sqrt{\frac{A+m^2}{-\frac{A}{3}+m^2}}\right]\right)+\frac{7}{9} A \ln\left[\frac{-\frac{A}{3}+m^2}{\overline \mu^2}\right]\;.
\end{eqnarray}
After determining these conditions, we still need to choose an appropriate value for $\overline \mu$. In order to compare possible other minima with the minimum \eqref{min1}, we should operate with the same scale. Therefore, we set
\begin{eqnarray}
\overline \mu^2 &\approx& 4.12 \lms^2 \;,
\end{eqnarray}
while with this choice, the coupling constant becomes,
\begin{eqnarray}
g^2(\overline{\mu}^2)&=& \frac{1}{\beta_0\ln\frac{\overline{\mu}^2}{\lms^2} } \nonumber\\
\Rightarrow \frac{g^2 N}{16 \pi^2}&=& \frac{3}{11 \ln (4.12)} \approx 0.19\;,
\end{eqnarray}
which is sufficiently small to assure a trustworthy perturbative
expansion, as we have carried out.

If one solves the two gap equations numerically, besides the minimum \eqref{min1}, one only finds the maximum
\begin{eqnarray}
A&=&0\;, \quad m^2 ~=  0\;.
\end{eqnarray}
In summary, the potential \eqref{potentiaal} has only one minimum \eqref{min1},
\begin{eqnarray}
V^{(1)}(m^2= 4.12, A = 0)&\approx&-3.23 \lms^4\;,
\end{eqnarray}
where we have set $N=3$. As $A=0$ in the minimum, Lorentz
invariance is preserved as required in this $T=0$ case. This is a
good test for the solidness of our framework, before going to the
more complicated case of finite temperature.

\section{Conclusion}
In this paper we have introduced an analytical formalism which allows one to study the electric-magnetic asymmetry of the dimension two condensate $\braket{A_\mu^2}$ first considered in \cite{Chernodub:2008kf} by means of lattice simulations.

The first main result is that such a formalism exists: it is possible to compute the expection value of $A_\mu^2$ and of its asymmetry simultaneously, generalizing the formalism developed in \cite{Verschelde:2001ia}. We do this in the Landau gauge, where $\int \d^dx\; A_\mu^2$ reaches its minimum when $A_\mu$ moves along its gauge orbit, giving a gauge invariant meaning to $\braket{A_\mu^2}$ in the Landau gauge. Adding terms quadratic in the sources makes the model renormalizable, and performing a Hubbard-Stratonovich transform eliminates these terms, introducing new fields but restoring the usual energy interpretation of the effective action in a simple fashion.

As a second result, we have computed the parameters which appear in the formalism. This part of the calculation is based on the renormalization group. Once these parameters were fixed, we have determined the effective action itself. At zero temperature it has the same minimum as the one already found in \cite{Verschelde:2001ia} without considering the electric-magnetic asymmetry: there is a nonzero value for $\braket{A_\mu^2}$, and the asymmetry is zero as must be the case based on Lorentz invariance. No other minima were found.

Now that a consistent formalism has been developed at zero temperature, we can consider the finite temperature case.\cite{future}

\section*{Acknowledgments}
We wish to thank M.~Chernodub for encouraging this research. D.~Dudal and N.~Vandersickel are grateful for the hospitality at the Center for Theoretical Physics (MIT), where parts of this work were done.

\appendix

\section{Higher loop calculations in most general form}\label{appA}
In this Appendix we record the more general forms of the various LCO quantities
associated with the traceless operator. In essence they contain the dependence
on $\Nf$ massless quarks. We have checked that in the $\Nf$~$\rightarrow$~$0$
limit they return to the Yang-Mills values we have used and recorded in the
main body of the article. Though for the first two quantities, $Z_K$ and
$\delta \omega$, we have also included the gauge parameter dependence. We have
\begin{eqnarray}
Z_K &=& 1 ~+~ \left[ \left( \frac{1}{2} \alpha - \frac{29}{6} \right) C_A
+ \frac{8}{3} T_F \Nf \right] \frac{g^2}{16\pi^2\epsilon} \nonumber \\
&& +~ \left[ \left( \left( \frac{2117}{72} - \frac{19}{6} \alpha
- \frac{1}{8} \alpha^2 \right) C_A^2 + \frac{64}{9} T_F^2 \Nf^2
+ \left( \frac{4}{3} \alpha - \frac{262}{9} \right) T_F \Nf C_A \right)
\frac{1}{\epsilon^2} \right. \nonumber \\
&& \left. ~~~~~~+~ \left( \left( \frac{1}{8} \alpha^2 + \frac{43}{48} \alpha
- \frac{389}{48} \right) C_A^2 + 4 T_F \Nf C_F + \frac{31}{6} T_F \Nf C_A
\right) \frac{1}{\epsilon} \right] \frac{g^4}{(16\pi^2)^2} \nonumber \\
&& +~ \left[ \left( 27 \alpha^3 + 423 \alpha^2 + 9033 \alpha - 82563 \right)
C_A^3 + \left( 115616 - 7440 \alpha - 144 \alpha^2 \right) T_F \Nf C_A^2
\right. \nonumber \\
&& \left. ~~~~~~+~ \left( 1536 \alpha - 53504 \right) T_F^2 \Nf^2 C_A
+ 8192 T_F^3 \Nf^3 \right) \frac{1}{432\epsilon^3} \nonumber \\
&& \left. ~~~~~~+~ \left( \left( 99627 - 11048 \alpha - 1005 \alpha^2
- 90 \alpha^3 \right) C_A^3 + \left( 288 \alpha^2 + 5464 \alpha
- 110488 \right) T_F \Nf C_A^2 \right. \right. \nonumber \\
&& \left. \left. ~~~~~~~~~~~~~+~ \left( 1728 \alpha - 44736 \right)
T_F \Nf C_F C_A + 30080 T_F^2 \Nf^2 C_A + 21504 T_F^2 \Nf^2 C_F \right)
\frac{1}{864\epsilon^2} \right. \nonumber \\
&& \left. ~~~~~~+~ \left( \left( 567 \alpha^3 + 486 \zeta(3) \alpha^2
+ 3258 \alpha^2 + 3024 \zeta(3) \alpha + 15750 \alpha + 2754 \zeta(3) - 196111
\right) C_A^3 \right. \right. \nonumber \\
&& \left. \left. ~~~~~~~~~~~~~+~ \left( 268672 - 81216 \zeta(3) - 5688 \alpha
\right) T_F \Nf C_A^2 + \left( 19152 + 103680 \zeta(3) \right) T_F \Nf C_F C_A
\right. \right. \nonumber \\
&& \left. \left. ~~~~~~~~~~~~~-~ 41152 T_F^2 \Nf^2 C_A - 10368 T_F \Nf C_F^2
- 25344 T_F^2 \Nf^2 C_F \right) \frac{1}{7776\epsilon} \right]
\frac{g^6}{(16\pi^2)^3} ~+~ O(g^8)
\end{eqnarray}
and
\begin{eqnarray}
\delta \omega &=& \frac{\NA}{16\pi^2} \left[ -~ \left( \frac{7}{12}
+ \frac{1}{3} \alpha + \frac{1}{12} \alpha^2 \right) \frac{1}{\epsilon}
\right. \nonumber \\
&& \left. ~~~~~~~~~~~+~ \left( \left( \left( \frac{203}{72}
+ \frac{43}{72} \alpha
+ \frac{1}{24} \alpha^2 + \frac{1}{24} \alpha^3 \right) C_A ~-~
\left( \frac{14}{9} + \frac{4}{9} \alpha \right) T_F \Nf \right)
\frac{1}{\epsilon^2} \right. \right. \nonumber \\
&& \left. \left. ~~~~~~~~~~~~~~~~~~~+~ \left( -~ \left( \frac{1345}{864}
+ \frac{287}{864} \alpha + \frac{43}{288} \alpha^2
+ \frac{1}{288} \alpha^3 \right) C_A ~+~ \left( \frac{41}{54}
+ \frac{2}{27} \alpha \right) T_F \Nf \right) \frac{1}{\epsilon} \right)
\frac{g^2}{16\pi^2} \right. \nonumber \\
&& \left. ~~~~~~~~~~~+~ \left( \left( \left( \frac{49}{3}
+ \frac{196}{81} \alpha + \frac{1}{27} \alpha^2 \right) C_A T_F \Nf ~-~
\left( \frac{112}{27} + \frac{64}{81} \alpha \right) T_F^2 \Nf^2
\right. \right. \right. \nonumber \\
&& \left. \left. \left. ~~~~~~~~~~~~~~~~~~-~ \left( \frac{3451}{216}
+ \frac{1025}{648} \alpha + \frac{25}{216} \alpha^2 + \frac{5}{72} \alpha^3
+ \frac{1}{36} \alpha^4 \right) C_A^2 \right) \frac{1}{\epsilon^3}
\right. \right. \nonumber \\
&& \left. \left. ~~~~~~~~~~~~~~~~~~~+~ \left( \left( \frac{164}{81}
+ \frac{32}{243} \alpha \right) T_F^2 \Nf^2 ~-~ \left( \frac{28}{9}
+ \frac{8}{9} \alpha \right) C_F T_F \Nf ~-~ \left( \frac{449}{36}
+ \frac{446}{243} \alpha + \frac{49}{324} \alpha^2 \right) C_A T_F \Nf
\right. \right. \right. \nonumber \\
&& \left. \left. \left. ~~~~~~~~~~~~~~~~~~+~ \left( \frac{39203}{2592}
+ \frac{16717}{7776} \alpha + \frac{793}{2592} \alpha^2
+ \frac{31}{288} \alpha^3 + \frac{7}{432} \alpha^4
\right) C_A^2 \right) \frac{1}{\epsilon^2} \right. \right. \nonumber \\
&& \left. \left. \left. ~~~~~~~~~~~~~~~~~~+~ \left( \left( \frac{235}{243}
+ \frac{160}{729} \alpha \right) T_F^2 \Nf^2 ~+~ \left( \frac{295}{54}
- \frac{64}{9} \zeta(3) + \frac{13}{9} \alpha - \frac{16}{9} \zeta(3) \alpha
\right) C_F T_F \Nf \right. \right. \right. \right. \nonumber\\
&& \left. \left. \left. \left. ~~~~~~~~~~~~~~~~~~~~~~~~~+~ \left(
\frac{80}{9} \zeta(3)
- \frac{3991}{1944} - \frac{703}{2916} \alpha + \frac{16}{9} \zeta(3) \alpha
+ \frac{149}{1944} \alpha^2 \right) C_A T_F \Nf \right. \right. \right. \right.
\nonumber
\\&& \left. \left. \left. \left. ~~~~~~~~~~~~~~~~~~~~~~~~~-~ \left(
\frac{9881}{3456} + \frac{52709}{46656} \alpha + \frac{407}{3888} \alpha^2
+ \frac{29}{576} \alpha^3 + \frac{65}{10368} \alpha^4
\right. \right. \right. \right. \right. \nonumber \\
&& \left. \left. \left. \left. \left. ~~~~~~~~~~~~~~~~~~~~~~~~~~~~~~~~~+~
\left( \frac{1481}{576}
+ \frac{19}{144} \alpha + \frac{55}{288} \alpha^2
+ \frac{1}{48} \alpha^3 + \frac{1}{576} \alpha^4 \right) \zeta(3)
\right) C_A^2 \right) \frac{1}{\epsilon} \right) \right)
\frac{g^4}{(16\pi^2)^2} \right] ~+~ O(g^6) ~.
\end{eqnarray}
Given these we find
\begin{eqnarray}
\gamma_K(a) &=& -~ \left[ ( 3 \alpha - 29 ) C_A + 16 T_F \Nf \right]
\frac{a}{6} \nonumber \\
&& ~~~~~+~ \left[ ( 6 \alpha^2 + 43 \alpha - 389 ) C_A^2 + 248 C_A T_F \Nf
+ 192 C_F T_F \Nf \right] \frac{a^2}{24} \nonumber \\
&& ~~~~~+~ \left[ ( 567 \alpha^3 + 486 \zeta(3) \alpha^2 + 3258 \alpha^2
+ 3024 \zeta(3) \alpha + 15750 \alpha + 2754 \zeta(3) - 196111 ) C_A^3
\right. \nonumber \\
&& \left. ~~~~~~~~~~+~ ( 268672 - 81216 \zeta(3) - 5688 \alpha ) C_A^2 T_F \Nf
+ ( 103680 \zeta(3) + 19152 ) C_A C_F T_F \Nf \right. \nonumber \\
&& \left. ~~~~~~~~~~-~ 41152 C_A T_F^2 \Nf^2 - 10368 C_F^2 T_F \Nf
- 25344 C_F T_F^2 \Nf^2 \right] \frac{a^3}{2592} ~+~ O(a^4)
\end{eqnarray}
and
\begin{eqnarray}
h(a) &=& -~ \left[ \alpha^2 + 4 \alpha + 7 \right] \frac{N_A}{12}
\nonumber \\
&& +~ \left[ ( 656 + 64 \alpha ) T_F \Nf - ( 3 \alpha^3 + 129 \alpha^2
+ 287 \alpha + 1345 ) C_A \right] \frac{N_A a}{432}
\nonumber \\
&& +~ \left[ ( 7152 \alpha^2 - 22496 \alpha + 165888 \zeta(3) \alpha
- 191568 + 829440 \zeta(3) ) C_A T_F \Nf \right. \nonumber \\
&& \left. ~~~~~-~ ( ( 162 \alpha^4 + 1944 \alpha^3 + 17820 \alpha^2
+ 12312 \alpha + 239922 ) \zeta(3) \right. \nonumber \\
&& \left. ~~~~~~~~~~+~ ( 585 \alpha^4 + 4698 \alpha^3 + 9768 \alpha^2
+ 105418 \alpha + 266787 ) ) C_A^2 \right. \nonumber \\
&& \left. ~~~~~+~ ( 134784 \alpha - 165888 \zeta(3) \alpha + 509760
- 663552 \zeta(3) ) C_F T_F \Nf \right. \nonumber \\
&& \left. ~~~~~+~ ( 20480 \alpha + 90240 ) \Nf^2 T_F^2 \right]
\frac{N_A a^2}{31104} ~+~ O(a^3)\;.
\end{eqnarray}
Finally, we find the Landau gauge version of $\omega$ for $\Nf$~$\neq$~$0$ is
\begin{eqnarray}
\omega &=& \frac{\NA}{16\pi^2} \left[
\frac{7}{4[ 7 C_A - 8 T_F \Nf ]} \right. \nonumber \\
&& \left. ~~~~~~~~~~+~ \left(  511 C_A^2 - 2452 C_A T_F \Nf + 1512 C_F T_F \Nf
+ 1312 T_F^2 \Nf^2 \right) \frac{g^2}{[ 29 C_A - 16 T_F \Nf ]
[ 7 C_A - 8 T_F \Nf ] 576 \pi^2} \right. \nonumber \\
&& \left. ~~~~~~~~~~+~ \left( ( 17352846 \zeta(3) - 10661959 ) C_A^4
+ ( 75444728 - 117227088 \zeta(3) ) C_A^3 T_F \Nf \right. \right. \nonumber \\
&& \left. \left. ~~~~~~~~~~~~~~~~~~
+~ ( 123538176 \zeta(3) - 83836800 ) C_A^2 T_F^2 \Nf^2
+ ( 86994432 \zeta(3) - 62449632 ) C_A^2 T_F \Nf C_F \right. \right.
\nonumber \\
&& \left. \left. ~~~~~~~~~~~~~~~~~~
-~ 2104704 C_A C_F^2 T_F \Nf
+ ( 59943168 - 99311616 \zeta(3) ) C_A C_F T_F^2 \Nf^2
\right. \right. \nonumber \\
&& \left. \left. ~~~~~~~~~~~~~~~~~~
+~ ( 32150528 - 35389440 \zeta(3) ) C_A T_F^3 \Nf^3
- 3850240 T_F^4 \Nf^4
\right. \right. \nonumber \\
&& \left. \left. ~~~~~~~~~~~~~~~~~~
+~ ( 28311552 \zeta(3) - 12865536 ) T_F^3 \Nf^3 C_F
+ 8128512 T_F^2 \Nf^2 C_F^2 \right) \right. \nonumber \\
&& \left. ~~~~~~~~~~~~~~~~~~
\times ~ \frac{g^4}{[ 29 C_A - 16 T_F \Nf ] [ 17 C_A - 8 T_F \Nf ]
[ 7 C_A - 8 T_F \Nf ] 2654208 \pi^4} \right] ~.
\end{eqnarray}
Hence we can deduce that
\begin{eqnarray}
Z_\omega &=& 1 ~+~ \frac{[11 C_A - 4 T_F \Nf ]g^2}{24\pi^2\epsilon}
\nonumber \\
&& +~ \left( 38857 C_A^3 - 34948 C_A^2 T_F \Nf - 22176 C_A C_F T_F \Nf
\right. \nonumber \\
&& \left. ~~~~+~ 6880 C_A T_F^2 \Nf^2 + 8064 C_F T_F^2 \Nf^2 \right)
\frac{g^4}{32256[29 C_A - 16 T_F \Nf] \pi^4 \epsilon} ~+~ O(g^6) ~.
\end{eqnarray}

\section{Details concerning the calculation of part III of the effective potential}
In this Appendix we shall calculate step by step part III of equation \eqref{start}, which is far from trivial.
\begin{eqnarray}
\mathrm{III} &=&\tr\ln\left(-\p^2+m^2 + A\left(1-\frac d{d-1}\frac{\p_0^2}{\p^2}\right)\right) \nonumber\\
&=& \int \frac{\d^d k}{(2\pi)^d} \ln\left(k^2 + m^2 + A\left(1-\frac d{d-1}\frac{k_0^2}{k^2}\right)\right) \nonumber\\
&=& \int \frac{\d^d k}{(2\pi)^d} \ln\left(k^4+k^2m^2 + A\left(k^2-\frac d{d-1} k_0^2\right)\right) - \tr\ln k^2 \nonumber \\
&=& \int \frac{\d^d k}{(2\pi)^d} \ln\left(k_0^4+k_0^2\left(m^2+2
k_i^2-\frac A{d-1}\right) + k_i^2(k_i^2+m^2+A)\right)\;,
\end{eqnarray}
where the notation $k_i$ refers to $(d-1)$-dimensional spatial part of $k$. In this case, notice that for III to be real valued, we must have
\begin{eqnarray}\label{voorwaarde22}
m^2+ A &\geq& 0 \;,
\end{eqnarray}
where we have assumed that expression \eqref{voorwaarde1} is certainly fulfilled.  We can split the integral in two parts, resulting in
\begin{eqnarray}\label{tssdoor}
\mathrm{III} &=& \int \frac{\d^{d-1} k_i}{(2\pi)^{d-1}} \int \frac{\d k_0}{2\pi}  \ln\left(k_0^2+\frac{m^2}2+k_i^2-\frac A{2(d-1)}+\sqrt{\frac{A^2}{4(d-1)^2}-\frac d{d-1}Ak_i^2+\frac{m^4}4-\frac{Am^2}{2(d-1)}}\right) \nonumber\\
&+ & \int \frac{\d^{d-1} k_i}{(2\pi)^{d-1}} \int \frac{\d k_0}{2\pi}
\ln\left(k_0^2+\frac{m^2}2+k_i^2-\frac
A{2(d-1)}-\sqrt{\frac{A^2}{4(d-1)^2}-\frac
d{d-1}Ak_i^2+\frac{m^4}4-\frac{Am^2}{2(d-1)}}\right) \;.
\end{eqnarray}
Next, we can perform the integration over $k_0$. In general, we can
write,
\begin{eqnarray}
\int \frac{\d^d k}{(2 \pi)^d} \ln (k^2 + x) &=& \int \frac{\d^{d-1}
k_i}{(2\pi)^{d-1}} \int \frac{\d k_0}{2 \pi} \ln(k_0^2 + k_i^2 + x)
-\int \frac{\d^{d-1} k_i}{(2\pi)^{d-1}} \int \frac{\d k_0}{2 \pi}
\ln(k_0^2 + k_i^2)\;,
\end{eqnarray}
as the second part is zero in dimensional regularization,
\eqref{logint}. Evaluating the integral over $k_0$ gives,
\begin{eqnarray}
\int \frac{\d^d k}{(2 \pi)^d} \ln(k^2 + x) &=& \int \frac{\d^{d-1}
k_i}{(2\pi)^{d-1}} \left( \sqrt{k_i^2 + x} - \sqrt{k_i^2}\right)\;.
\end{eqnarray}
The second part is again zero in dimensional regularization so we obtain the following general formula,
\begin{eqnarray}
\int \frac{\d^d k}{(2 \pi)^d} \ln(k^2 + x) &=& \int \frac{\d^{d-1}
k_i}{(2\pi)^{d-1}}\sqrt{k_i^2 + x}  \;,
\end{eqnarray}
which we can apply to expression \eqref{tssdoor}
\begin{eqnarray}
\mathrm{III} &=& \int \frac{\d^{d-1} k_i}{(2\pi)^{d-1}} \sqrt{\frac{m^2}2+k_i^2-\frac{A}{2(d-1)}+\sqrt{\frac{A^2}{4(d-1)^2}-\frac d{d-1}A k_i^2+\frac{m^4}4-\frac{Am^2}{2(d-1)}}} \nonumber\\
&& + \int \frac{\d^{d-1} k_i}{(2\pi)^{d-1}}
\sqrt{\frac{m^2}2+k_i^2-\frac
A{2(d-1)}-\sqrt{\frac{A^2}{4(d-1)^2}-\frac
d{d-1}Ak_i^2+\frac{m^4}4-\frac{Am^2}{2(d-1)}}}\;.
\end{eqnarray}
The next step will be to simplify this expression. If we define:
\begin{eqnarray}
a&=& \frac{m^2}2+k_i^2-\frac{A}{2(d-1)}\;,\nonumber\\
b&=&\sqrt{\frac{A^2}{4(d-1)^2}-\frac d{d-1}A k_i^2+\frac{m^4}4-\frac{Am^2}{2(d-1)}}\;,
\end{eqnarray}
we need to simplify the following expression,
\begin{eqnarray}\label{a+b}
\sqrt{a+b} + \sqrt{a-b}\;.
\end{eqnarray}
Notice that due to the constraint \eqref{voorwaarde1}, $a > 0$ for
all values of $k_i^2$. Let us first assume $A\geq0$. For
sufficiently small $k_i^2$, $b$ will be a positive real number,
smaller than $a$. We may write
\begin{eqnarray}\label{a+b2}
\left((a+b)^{1/2} + (a-b)^{1/2} \right)^2 &=& \left((a+ b)^{1/2} \right)^{2} + \left((a- b)^{1/2} \right)^{2} + 2(a+ b)^{1/2} (a - b)^{1/2}  \nonumber\\
&=& 2a + 2 (a^2- b^2)^{1/2}\;,
\end{eqnarray}
and taking the square of this equation results in
\begin{eqnarray}\label{a+b3}
\sqrt{a+b}+\sqrt{a-b}=\sqrt2\sqrt{a+\sqrt{a^2-b^2}}\;.
\end{eqnarray}
For $k_i^2$ larger than a certain value, the argument of the square
root defining $b$ will flip sign and $b$ will become purely
imaginary, i.e.~$b = i b'$ with $b'
> 0$. The derivations \eqref{a+b2} and \eqref{a+b3} remain valid,
keeping in mind that $(a + ib')$ always lies in the first quadrant
of the complex plane, and $(a-ib')$ in the fourth quadrant.

For $A<0$, we necessarily have that $a\geq b>0$, which can be easily
checked using the constraints \eqref{voorwaarde1} and
\eqref{voorwaarde2}. Also now, \eqref{a+b2} and \eqref{a+b3} go
through, and we conclude that, given the original conditions
\eqref{voorwaarde1} and \eqref{voorwaarde2}, we can always employ
the equality in \eqref{a+b3}.

Using this formula, we can rewrite the integral III in the following form
\begin{eqnarray}
\mathrm{III}&=& \sqrt2 \int \frac{\d^{d-1} k_i}{(2\pi)^{d-1}}
\sqrt{\frac{m^2}2+k_i^2-\frac
A{2(d-1)}+\sqrt{k_i^2}\sqrt{k_i^2+A+m^2}}\;.
\end{eqnarray}
With the help of the following adapted Schwinger trick
\begin{eqnarray}
\sqrt{A+\ell \sqrt B} &=& -\frac \ell {4\pi}\lim_{z\to-\frac{1}{2}}
\int_0^\infty \d t t^z \int_0^\infty \frac{\d s}{s^{3/2}}
\e^{-\frac{t^2\ell^2}{4s}-tA-sB}\;,
\end{eqnarray}
we can rewrite the square root
\begin{eqnarray}\label{eq2}
\mathrm{III} &=& - \frac{1}{\sqrt8\pi} \int \frac{\d^{d-1} k_i}{(2\pi)^{d-1}}\left(  \lim_{z\to-\frac{1}{2}} \int_0^\infty \d t t^z \int_0^\infty \frac{\d s}{s^{3/2}} \sqrt{k_i^2} \e^{-(\frac{t^2}{4s}+t+s) k_i^2} e^{-t(\frac{m^2}2-\frac A{2(d-1)})-s(A+m^2)} \right) \nonumber\\
&=&  - \frac{1}{\sqrt8\pi} \lim_{z\to-\frac{1}{2}}  \int_0^\infty \d t t^z \int_0^\infty \frac{\d s}{s^{3/2}}  e^{-t(\frac{m^2}2-\frac A{2(d-1)})-s(A+m^2)} \left(\int \frac{\d^{d-1} k_i}{(2\pi)^{d-1}} \sqrt{k_i^2} \e^{-(\frac{t^2}{4s}+t+s) k_i^2} \right)\;.
\end{eqnarray}
Therefore, we can now evaluate the integral over $k$, yielding
\begin{eqnarray}\label{eq1}
\int \frac{\d^{d-1} k}{(2\pi)^{d-1}} k \e^{-(\frac{t^2}{4s}+t+s) k^2} &=& \frac{V_{d-1}}{(2\pi)^{d-1}} \int_0^\infty k^{d-1} \e^{-(\frac{t^2}{4s}+t+s) k^2}  \d k \nonumber\\
&=& \frac1{2^{d-1}\pi^{(d-1)/2}}
\left(\frac{t^2}{4s}+t+s\right)^{-d/2} \frac{\Gamma(\frac d2)}{\Gamma(\frac{d-1}2)}\;.
\end{eqnarray}
If we insert equation \eqref{eq1} into equation \eqref{eq2}, we obtain,
\begin{eqnarray}
\mathrm{III} &=& -\frac1{\sqrt2\pi^{(d+1)/2}} \frac{\Gamma(\frac d2)}{\Gamma(\frac{d-1}2)} \quad \lim_{z\to-\frac{1}{2}} \quad \int_0^\infty \d t t^z \int_0^\infty \d s \frac{s^{(d-3)/2}\e^{-t(\frac{m^2}2-\frac A{2(d-1)})} \e^{-s(A+m^2)}}{(t+2s)^d} \nonumber\\
&=& -\frac1{2^{d/2}\pi^{(d+1)/2}} \frac{\Gamma(\frac d2)}{\Gamma(\frac{d-1}2)} \quad \lim_{z\to-\frac{1}{2}} \quad \int_0^\infty
\frac{\d t}{t^{(d-1)/2-z}}\e^{-t(\frac{m^2}2-\frac A{2(d-1)})}
\int_0^\infty \d s' \e^{-s't(A+m^2)/2} s^{\prime \ (d-3)/2} (1+s')^{-d}\;,
\end{eqnarray}
where in the last step, we have performed the substitution $s' = 2 s/t$. In this expression, we can switch the integral and the limit as it will turn out that the integral will converge (within the constraints \eqref{voorwaarde1} and \eqref{voorwaarde2}), yielding,
\begin{eqnarray}\label{tss2}
\mathrm{III} &=& -\frac1{2^{d/2}\pi^{(d+1)/2}} \frac{\Gamma(\frac d2)}{\Gamma(\frac{d-1}2)}  \quad \int_0^\infty
\frac{\d t}{t^{d/2+1}}\e^{-t(\frac{m^2}2-\frac A{2(d-1)})}
\int_0^\infty \d s' \e^{-s't(A+m^2)/2} s^{\prime \ (d-3)/2} (1+s')^{-d}\;.
\end{eqnarray}
We recognize a Kummer function of the second kind,
\begin{eqnarray}
U(a,b,z) &=& \frac{1}{\Gamma(a)} \int_0^{\infty}\d x \e^{-zx} x^{a-1} (1+x)^{b-a-1} \nonumber\\
&=& \frac{\pi}{ \sin (\pi b)} \left[ \frac{_1 F_1(a;b;z) }{\Gamma(a-b+1) \Gamma(b)} - \left(z^{1-b}\right) \frac{  _1F_1(a - b + 1;2-b;z) }{\Gamma(a) \Gamma(2-b)}\right]\;,
\end{eqnarray}
where is $_1 F_1(a;b;z)$ a confluent hypergeometric function of the first kind. Comparing this Kummer function with expression \eqref{tss2} we can write,
\begin{eqnarray}
\mathrm{III}&=&-\frac{\Gamma(\frac{d}{2})}{2^{d/2}\pi^{(d+1)/2}}    \int_0^\infty \frac{\d t}{t^{d/2+1}}\e^{-t(\frac{m^2}2-\frac A{2(d-1)})} U\left(\frac{d-1}2,-\frac{d-1}2,\frac{t}{2}(A+m^2)\right) \nonumber\\
&=&\frac{\Gamma(\frac d2)}{2^{d/2}\pi^{(d-1)/2}\sin\pi\frac{d-1}2} \int_0^\infty \frac{dt}{t^{d/2+1}}\e^{-t(\frac{m^2}2-\frac{A}{2(d-1)})} \Biggl(\frac{_1F_1(\frac{d-1}2;-\frac{d-1}2;\frac
{t}{2}(A+m^2))}{\Gamma(d)\Gamma(-\frac{d-1}2)} \nonumber\\
&& \hspace{8cm}- \left(\frac{t}{2}(A+m^2)\right)^{\frac{d+1}{2}} \frac{_1F_1\left(d;\frac{d+3}2;\frac
t2(A+m^2)\right)}{\Gamma(\frac{d-1}2)\Gamma(\frac{d+3}2)} \Biggr)\;.
\end{eqnarray}
For the final integration, we recognize again a hypergeometric function,
\begin{eqnarray}
_2 F_1(a,b;c;z) &=& \frac{1}{\Gamma(b)} \int_0^{+\infty} \d t \e^{-t} t^{b-1} \ _1F_1(a,c, tz)\;,
\end{eqnarray}
resulting in
\begin{equation}\label{III}
\mathrm{III} = \frac{\Gamma(\frac d2)}{2^{d/2}\pi^{(d-1)/2}\sin\pi\frac{d-1}2}
\left(\frac{_2F_1\left(\frac{d-1}2,-\frac
d2;-\frac{d-1}2;\frac{\frac12(A+m^2)}{(\frac{m^2}2-\frac
A{2(d-1)})}\right)\Gamma(-\frac d2)}{\left(\frac{m^2}2-\frac
A{2(d-1)}\right)^{-d/2}\Gamma(d)\Gamma(-\frac{d-1}2)} -
\left(\frac{A+m^2}2\right)^{\frac{d+1}2}
\frac{_2F_1\left(d,\frac12;\frac{d+3}2;\frac{\frac12(A+m^2)}{(\frac{m^2}2-\frac
A{2(d-1)})}\right)\Gamma(\frac12)}{\left(\frac{m^2}2-\frac
A{2(d-1)}\right)^{1/2}\Gamma(\frac{d-1}2)\Gamma(\frac{d+3}2)} \right)\;,
\end{equation}
where this expression is defined in $d$ dimensions.
The next step is to replace $d \to 4 - \epsilon$, and to rewrite
this expression in a series in $\epsilon$. Only in the first term,
we do encounter a pole $1/\epsilon$ originating from
$\Gamma(-\frac{d}{2})$. In the second term, there is no such pole
and, therefore, we can immediately set $d = 4$:
\begin{eqnarray}
\mathrm{III}_b = \left. \frac{-\Gamma\left(\frac d2\right)\left(\frac{A+m^2}2\right)^{\frac{d+1}2}}{2^{d/2}\pi^{(d-1)/2}\sin\pi\frac{d-1}2} \;
\frac{_2F_1\left(d,\frac12;\frac{d+3}2;\frac{\frac12(A+m^2)}{(\frac{m^2}2-\frac
A{2(d-1)})}\right)\Gamma\left(\frac{1}{2}\right)}{\left(\frac{m^2}2-\frac{A}{2(d-1)}\right)^{1/2}\Gamma(\frac{d-1}2)\Gamma(\frac{d+3}2)}\right|_{d=4}
&=&  \frac{ (m^2+A)^{5/2}}{15\pi^2\sqrt{m^2-\frac{A}{3}}}{}_2F_1\left(4,\frac{1}{2};\frac{7}{2};\frac{m^2+A}{m^2-\frac{A}{3}}\right)\;.
\end{eqnarray}
For the first term, we have to expand in a series of
$\epsilon$. For the benefit of the reader, we shall do this
expansion in a structured way. We can distinguish 3 different parts.
A pre-factor, the hypergeometric function and
$\Gamma(-\frac{d}{2})$. Firstly, after some algebra, we can write
the expanded pre-factor as,
\begin{eqnarray}
&&\frac{\Gamma(\frac d2)}{2^{d/2}\pi^{(d-1)/2}\sin\pi\frac{d-1}2} \frac{1}{\left(\frac{m^2}2-\frac{A}{2(d-1)}\right)^{-d/2}\Gamma(d)\Gamma(-\frac{d-1}2)} \nonumber\\
&=& \frac{-1}{8(4\pi)^2} \left\{\left(  m^2 -\frac{A}{3}  \right)^2
- \frac{\epsilon}{2} \left(m^2 - \frac{A}{3} \right) \left[
\frac{4}{9} A + \left(m^2 - \frac{A}{3} \right) \left( \ln\left(
\frac{m^2 - \frac{A}{3}}{\mu^2}\right) - \ln (16 \pi) \right)
\right] \right\}\;,
\end{eqnarray}
where we are working in the MS-scheme (later we shall convert $\mu$ to the $\MSbar$ scheme). Let us mention that in the calculation of this expansion, we have encountered the digamma function of $-3/2$, i.e.~$\psi(-3/2)$, which can be reduced to
\begin{eqnarray}
\psi(-3/2) &=& -2 \ln 2 - \gamma + 8/3\;,
\end{eqnarray}
with the help of the following relations,
\begin{eqnarray}
\psi(z) &=& \psi(z + 1 ) - \frac{1}{z}\;,\nonumber\\
\psi\left(\frac{1}{2}\right)&=& -2 \ln 2 - \gamma\;.
\end{eqnarray}
Secondly, we have to expand the hypergeometric function into a
series in $\epsilon$. We find
\begin{eqnarray}\label{hypexp}
\hspace{-6mm}{}_2F_1\left(\frac{d-1}2,-\frac
d2;-\frac{d-1}2;\frac{\frac12(A+m^2)}{(\frac{m^2}2-\frac
A{2(d-1)})}\right)
&=& (1+2v+5v^2)\nonumber\\&+&\hspace{-1mm}\left(\frac{11v^3-3v}{3(1-v)}+\frac{2}{9} (1+5v) \frac{A v}{m^2 -
\frac{A}{3}}- \frac{1}{2} (1 + 2 v + 5 v^2) \ln(1-v)\right)\epsilon+\mathcal{O}(\epsilon^2)\;,
\end{eqnarray}
where
\begin{eqnarray}\label{nuschaal}
v&=& \frac{m^2 + A}{m^2 - \frac{A}{3}}\;.
\end{eqnarray}
We checked the explicit result for the expansion \eqref{hypexp} with the {\sc  Mathematica} package {\sc HypExp} \cite{Huber:2005yg,Huber:2007dx}. Finally, the expansion $\Gamma(-\frac{d}{2})$ in terms of
$\epsilon$ reads,
\begin{eqnarray}
\Gamma(-\frac{d}{2})  &=& \frac{1}{\epsilon} + \left( \frac{3}{4}
- \frac{\gamma}{2} \right) + O(\epsilon)\;.
\end{eqnarray}
Taking the three previous expansions in $\epsilon$ together, we
find,
\begin{eqnarray}
\mathrm{III}_a &=& -\frac{1}{18(4\pi)^2} \left[5 A^2 + 12 A m^2 + 9 m^4 \right]\left[ \frac{2}{\epsilon} - \ln\left( \frac{A}{A-3m^2}\right) - \ln\left( \frac{m^2- A/3}{\overline{\mu}^2}\right) \right] \nonumber\\
&& \hspace{6cm} + \frac{1}{108(4\pi)^2} \left( -7 A^2 + 15 A m^2 + 27 m^4 + 27 \frac{m^6}{A}\right)\;.
\end{eqnarray}
Notice that we have switched to the $\MSbar$ scheme. In summary, equation \eqref{III} becomes,
\begin{eqnarray}\label{partIII}
\mathrm{III} &=&
-\frac{1}{18(4\pi)^2} \left[5 A^2 + 12 A m^2 + 9 m^4 \right]\left[ \frac{2}{\epsilon} - \ln\left( \frac{A}{A-3m^2}\right) - \ln\left( \frac{m^2- A/3}{\overline{\mu}^2}\right) \right]\nonumber\\
&&  + \frac{1}{108(4\pi)^2} \left( -7 A^2 + 15 A m^2 + 27 m^4 + 27 \frac{m^6}{A}\right)  +\frac{ (m^2+A)^{5/2}}{15\pi^2\sqrt{m^2-\frac{A}{3}}}{}_2F_1\left(4,\frac{1}{2};\frac{7}{2};\frac{m^2+A}{m^2-\frac{A}{3}}\right)\;.
\end{eqnarray}

\end{document}